%% file: prd_tri_submit.tex
\RequirePackage{lineno}
\documentclass[aps,prd,twocolumn,showpacs,superscriptaddress,groupedaddress]{revtex4}  
\usepackage{graphicx}  
\usepackage{dcolumn}   
\usepackage{bm}        
\usepackage{amssymb}   
\usepackage{longtable}

\newcommand{\pz}{\phantom{0}}

\def\ppbar{{p}\bar{{p}}}

\def\ttbar{{t}\bar{{t}}}
\def\bbbar{{b}\bar{{b}}}
\def\METnoSpace{{\mbox{$E\kern-0.57em\raise0.19ex\hbox{/}_{T}$}}}
\def\MET{{\mbox{$E\kern-0.57em\raise0.19ex\hbox{/}_{T}$}}}
\def\METsig{{\mbox{${\cal S}(E\kern-0.57em\raise0.19ex\hbox{/}_{T})$}}}
\def\METspecial{{\mbox{$\hat{E}\kern-0.57em\raise0.19ex\hbox{/}_{T}$}}}

\begin{document}

\hspace{5.2in} \mbox{Fermilab-Pub-13-050-E}

\title{Search for Higgs boson production in trilepton and like-charge electron-muon
final states with the D0 detector}
\input author_list.tex

\date{February 21, 2013}

\begin{abstract}
We present a search for Higgs bosons 
in multilepton final states in $p\bar{p}$ collisions at $\sqrt{s}=1.96$~TeV
recorded with the D0 detector at the Fermilab Tevatron Collider, using the
full Run~II data set with integrated luminosities of up to $9.7$~fb$^{-1}$.
The multilepton states considered are $ee\mu$, $e\mu\mu$, $\mu\tau_h\tau_h$ and
like-charge $e^{\pm}\mu^{\pm}$ pairs. These channels directly probe
the ${\sl HVV}$ ($V=W,Z$) coupling of the Higgs boson in production 
and decay. The  $\mu\tau_h\tau_h$ channel is also sensitive to $H\to\tau^+\tau^-$ decays.
Upper limits at the $95\%$ C.L~on the rate of standard model Higgs boson production 
are derived in the mass range $100 \le M_H \le 200$~GeV. 
The expected and observed limits are a factor of 6.3 and 8.4 above the 
predicted standard model cross section at $M_H=125$~GeV. 
We also interpret the data in a fermiophobic Higgs boson model.

\end{abstract}

\maketitle
\section{Introduction}
The Higgs boson is predicted by the standard model (SM) as a consequence of the 
breaking of the electroweak symmetry, which gives mass to the weak gauge bosons. 
The ATLAS and CMS Collaborations at the CERN Large Hadron Collider 
have recently reported the observation of a Higgs-like boson 
at a mass of $M_H\approx125$~GeV, primarily in
$\gamma\gamma$ and $ZZ$ final states~\cite{bib-atlas,bib-cms}.
Combining searches in the channel where the Higgs boson is produced
in association with a  $W$ or $Z$ boson, the CDF and D0 Collaborations
have found evidence for Higgs boson decay into $\bbbar$ pairs~\cite{bib-evidence}.

In this Article, we study final states with multiple charged leptons, 
including electrons, muons, and hadronically decaying tau leptons ($\tau_h$). 
We present the first Higgs boson search performed in the
trilepton final states $ee\mu$, $e\mu\mu$, and $\mu\tau_h\tau_h$ with the 
D0 detector. 

We also consider the production of like-charge $e^{\pm}\mu^{\pm}$ pairs.
This final state has the advantage of reduced background from $Z$ boson decay 
that is present in opposite-charge $e^+e^-$, $\mu^+\mu^-$, and $e^{\pm}\mu^{\mp}$ 
final states~\cite{dilepton}.
This analysis supersedes the previous searches in 
$e^{\pm}\mu^{\pm}$ final states, which used
 integrated luminosities of up to $5.3$~fb$^{-1}$~\cite{bib-d0like2}.

The main Higgs boson production mechanisms relevant for this analysis
are associated {\sl WH} and {\sl ZH} production and gluon-gluon fusion.
The contribution from vector boson fusion is small and therefore neglected.
The multilepton channels are sensitive to Higgs boson decays 
into $W^+W^-$ and $ZZ$ pairs, where the vector bosons ($V$) decay leptonically.
These channels therefore directly probe the {\it HVV} coupling in production and decay.
The trilepton searches are also sensitive to $H\to\tau^+\tau^-$ decays
from associated production ({\it \hspace{-2mm} WH, ZH}) through hadronic tau decays
in the $\mu\tau_h\tau_h$ channel and through leptonic tau decays in
the $ee\mu$ and $e\mu\mu$ channels.
Searches for $H\to\tau^+\tau^-$ decays in final states
with additional jets have also been performed using the full Run~II data set~\cite{Htau}.

We also interpret the data in a fermiophobic model with a Higgs boson
that does not couple to fermions and couples to
$W$ and $Z$ bosons with SM strengths.
Such searches have been conducted at 
the CERN $e^+e^-$ Collider 
LEP~\cite{Abdallah:2003xf, Heister:2002ub,Acciarri:2000zp,Abbiendi:2002yc} and
at the Fermilab Tevatron Collider~\cite{Abazov:2011ix,Aaltonen:2009ga}.
The CMS Collaboration excludes
fermiophobic Higgs bosons with $M_H<124.5$~GeV, 
$127< M_H<147.5$~GeV, and $155<M_H<180$~GeV at the $95\%$~C.L.~in a model that
assumes the couplings of the Higgs boson to other 
bosons are SM-like~\cite{CMS:2012bd}. 

Most results in this Article are based on the full Run~II data set 
collected with the D0 detector at the Fermilab Tevatron
and correspond to an integrated luminosity of $9.7$~fb$^{-1}$.
The analysis of the $\mu\tau_h\tau_h$ final state only uses data
recorded after June 2006 with an integrated luminosity of $8.6$~fb$^{-1}$. 
The results provide an important input for the combined
Higgs boson search performed by the D0 Collaboration~\cite{bib-newcombo1}
and for the Tevatron combination~\cite{bib-newcombo2}.

\section{D0 Detector}
 
The D0 detector~\cite{d0det} comprises tracking detectors and calorimeters.
Silicon microstrip detectors and a
scintillating fiber tracker are used to reconstruct charged particle tracks 
within a $2$~T solenoid.    
A liquid-argon and uranium calorimeter has a 
central section (CC) covering pseudorapidities~\cite{eta} $|\eta_d|$ up to 
$\approx 1.1$, and two end calorimeters (EC) that extend coverage 
to $|\eta_d|\approx 4.2$. The calorimeters consist of electromagnetic
(EM) and hadronic sections segmented longitudinally in several layers.
Muons are identified by combining tracks with patterns of hits in the muon spectrometer, which
lies outside the calorimeter and
consists of a layer of tracking detectors and scintillation trigger 
counters in front of a 1.8~T toroid, followed by two similar layers 
after the toroid~\cite{muondetpaper}.
Trigger decisions are based on partial information from the
tracking detectors, calorimeters, and muon spectrometer.

\section{Event Simulation}

All background processes are simulated using Monte Carlo (MC) event generators, 
except the $Z\gamma$ background in
the $e\mu\mu$ channel and the multijet background, which are determined from data. 
The $W$+jets, $Z/\gamma^{*}\to {\ell^+\ell^-}$+jet, and $t\bar{t}$ processes 
are generated using {\sc alpgen}~\cite{bib-alpgen} with showering and
hadronization provided by {\sc pythia}~\cite{bib-pythia}. 
Diboson production ({\it \hspace{-2mm} WW}, {\it WZ}, and {\it ZZ}) and 
signal events are simulated using {\sc pythia}.
All these simulations use the CTEQ6L1~\cite{cteq6} parton distribution functions (PDFs).
Associated production of Higgs bosons ({\it \hspace{-2mm} WH} and {\it ZH}) and gluon-gluon fusion are generated in $5$~GeV increments of $M_H$ in the range 
$100\le M_H \le 200$~GeV. 
Tau lepton decays are simulated with {\sc tauola}~\cite{bib-tauola}, 
which includes a full treatment of the tau polarization.

Next-to-leading order (NLO) quantum chromodynamics calculations 
of cross sections are 
used to normalize the background contribution of $t \bar{t}$~\cite{ttcalc}
and diboson~\cite{mcfm} processes.
The $WZ$ production cross section is corrected for $W \gamma^*$ 
interference using {\sc powheg}~\cite{powheg}. 
The cross section for $W/Z$+jets production is normalized to a next-to-NLO (NNLO) 
calculation~\cite{ZWcalc}. 
The transverse momentum ($p_T$) spectrum of $Z$ bosons is corrected
to match the measured distributions~\cite{zptreweighting}. 
The correction factor for the $p_T$ spectrum of $W$ bosons is the product of
the $Z$ boson correction factor  and the ratio of the 
$p_T$ spectra of $W$ and $Z$ bosons calculated at NNLO~\cite{wptreweighting}.

The cross sections for {\sl VH} associated production are calculated at 
NNLO~\cite{higgsxec2,higgsxec3}.  The NNLO calculation
of Higgs boson production in gluon-gluon fusion takes into account
resummation of soft gluons to next-to-next-to-leading-log (NNLL)~\cite{higgsxsec1}. 
Higher order corrections to the Higgs boson production cross sections
are computed with the MSTW 2008 PDF set~\cite{mstw}.
The simulated $p_T$ spectrum of Higgs bosons from gluon-gluon fusion is corrected 
using the NNLO and NNLL calculation of {\sc hqt}~\cite{hqt}.  
Branching fractions of the Higgs boson decays are calculated using {\sc hdecay} \cite{hdecay}.

All MC samples are processed through a 
{\sc geant}~\cite{bib-geant} simulation of the detector. Data 
from random beam crossings are overlaid on the
MC events to account for detector noise and additional $\ppbar$ interactions.
The simulated distributions are corrected for differences between data and 
simulation in the reconstruction efficiencies and in the distribution of the longitudinal
coordinate of the interaction point.

To maximize signal acceptance, we use 
all events that pass our event selection without requiring 
a specific trigger condition. The residual efficiency loss from events
that are not recorded depends on the event kinematics.
We study ratios of kinematic distributions using the inclusive
trigger requirements and using a set of specific single lepton triggers. These ratios
are then used to derive corrections on the normalization and shape of the kinematic 
distributions for MC events.

\section{Object Identification}

The signal comprises electrons, muons, and tau leptons that are isolated from
other particles in the detector.

Electrons are characterized by their interaction and the resulting
shower shape in the EM calorimeter.
The electron clusters in the EM calorimeter are required to match a track in the central tracker. 
The energy is measured in the EM and 
the first hadronic layers of the calorimeter within a cone of radius
$\mathcal{R} = \sqrt{(\Delta\eta)^2 + (\Delta\phi)^2}$ = 0.2, where $\phi$ is the azimuthal angle.
The electron cluster must satisfy a set of criteria: 
(i) calorimeter isolation fraction, $f_{\rm iso} = (E_{\rm tot} - E_{\rm EM})/E_{\rm EM}$, 
less than 0.15 for the CC region and less than 0.1 for EC, 
where $E_{\rm tot}$ is the total energy in the cone 
of radius $\mathcal{R} = 0.4$ and $E_{\rm EM}$ is the EM energy in a cone of 
radius $\mathcal{R} = 0.2$;  
(ii) fraction of the EM energy to the total energy
greater than 0.9, and (iii) ratio of the electron's transverse momentum
measured by the calorimeter and by the tracking detector, respectively, 
less than 8 (CC only).
In addition, the value of an eight-variable likelihood
for electron candidates is required to be $\mathcal{L}_{e}>0.05$~\cite{bib-d0like2}.
The electrons must also satisfy a requirement on a neural network discriminant with 
seven input variables in the CC and three in the EC region, including
isolation and shower shape variables to improve the discrimination between 
jets and electrons.
The sum of the charged particle tracks' $p_T$ in an annulus of $0.05 < \mathcal{R} < 0.4$ around
the electron direction must be less than $3.5$~GeV in the CC and less than 
$(-2.5 \times |\eta_d| + 7)$~GeV in the EC.  

\begin{table*}
\caption{\protect\label{tab:cutflow}
Numbers of events in data, predicted background, and expected 
signal for $M_H=125$~GeV after the event selection.
 The numbers are shown for the different samples separately, together
 with their total (statistical and systematic) uncertainties.}
\renewcommand{\arraystretch}{1.0}
 \begin{center}
\begin{tabular}{l|c|c|c|c|c|c} 
\hline
\hline
&&&&&&\\ 
& $ee\mu$ & $e\mu\mu_{\rm A}$ & $e\mu\mu_{\rm B}$ & $e\mu\mu_{\rm C}$ & $\mu\tau_h\tau_h$ & 
 $e^{\pm}\mu^{\pm}$  \\
 \hline
Signal  &&&&&&\\
{\sl WH}          & $0.39$  & $0.23$ &  $0.07$  & $0.02$ & $0.55$  &  $1.93$\\
{\sl ZH}           & $0.45$  & $0.06$ &  $0.19$  & $0.11$ & $0.17$  &  $0.32$ \\
$gg\to H\to ZZ$      & $0.05$   & $<0.01$ &  $0.01$  & $0.02$ & $<0.01$ & $<0.01$ \\
\hline
Signal Sum   & $0.89$   & $0.29$  & $0.27$   & $0.15$ & $0.72$   & $2.25$\\
\hline
Background &&&&&&\\
$Z\to e^+e^-$  & $39.1 \pm 12.8$ & $<0.1$        & $<0.1$         & $<0.1$          & $0.3 \pm 0.1$  
&  $15.9  \pm 2.4\pz $\\
$Z\to \mu^+\mu^-$      & $<0.1$        & $\pz 2.6 \pm 0.9$ & $\pz 8.4 \pm 2.9$   & $32.3 \pm 10.6$  & $4.4 \pm 0.6$  
 & $58.5 \pm 15.2$\\
$Z\to \tau^+\tau^-$    & $1.5 \pm 0.6$  & $\pz 0.1 \pm 0.1$ & $\pz 0.1 \pm 0.1$   & $<0.1$          & $5.6 \pm 0.7$  & $22.0  \pm 6.8\pz$\\
$Z\gamma$          & $<0.1$         & $11.8 \pm 1.1$ & $24.2 \pm 2.0$  & $76.9 \pm 5.9\pz$       
& $<0.1$ & $<0.1$        \\
Diboson      & $37.1 \pm 4.3\pz$ & $\pz 3.9 \pm 0.5$ & $19.4 \pm 2.5$  & $9.4 \pm 1.2$   & $9.0 \pm 1.3$ & $36.2 \pm 3.6\pz$\\
$\ttbar$           & $1.2 \pm 0.2$  & $\pz 0.5 \pm 0.1$ & $\pz 0.3 \pm 0.1$   & $0.1 \pm 0.1$    & $1.4 \pm 0.2$  
& $4.1  \pm 2.1$\\
$W$+jets           & $0.2 \pm 0.1$   & $<0.1$        & $<0.1$         & $<0.1$          & $5.4 \pm 0.7$  
 & $238.3 \pm 19.0\pz$ \\
Multijet           & $<0.1$         & $<0.1$        & $<0.1$         & $<0.1$          & $<0.1$ &
 $434.5  \pm 87.0\pz$\\    
\hline
Background &&&&&\\
Sum           & $79 \pm 15$ & $19 \pm 2$ & $52 \pm 5$  & $119 \pm 11$ & $26 \pm 4$ & 
$809 \pm 93\pz$\\ \hline
Data               & $77$            & $16$           & $57$            & $119$            & $22$      & $822$ \\ 
\hline \hline
\end{tabular}
\end{center}
 \renewcommand{\arraystretch}{1.0}
 \end{table*}

Muons are identified by the presence of at least one track segment reconstructed 
in the muon spectrometer which is spatially consistent with a track in the central detector,
where the momentum and charge are measured by the curvature of this track.
Muon isolation is imposed with two isolation variables defined as the scalar sums of 
the transverse energy in the calorimeter in an annulus of radius 
$0.1 < \mathcal{R} < 0.4$ 
around the muon direction and of the momenta of charged particle tracks within 
$\mathcal{R} = 0.5$.
Both variables, divided by the muon $p_T$, must be less than $0.2$. 
To reduce the effects of charge mis-reconstruction,
additional selection criteria on the track quality are 
applied in the $e^{\pm}\mu^{\pm}$ channel.

\begin{figure*}[hbt]
\includegraphics[scale=0.29]{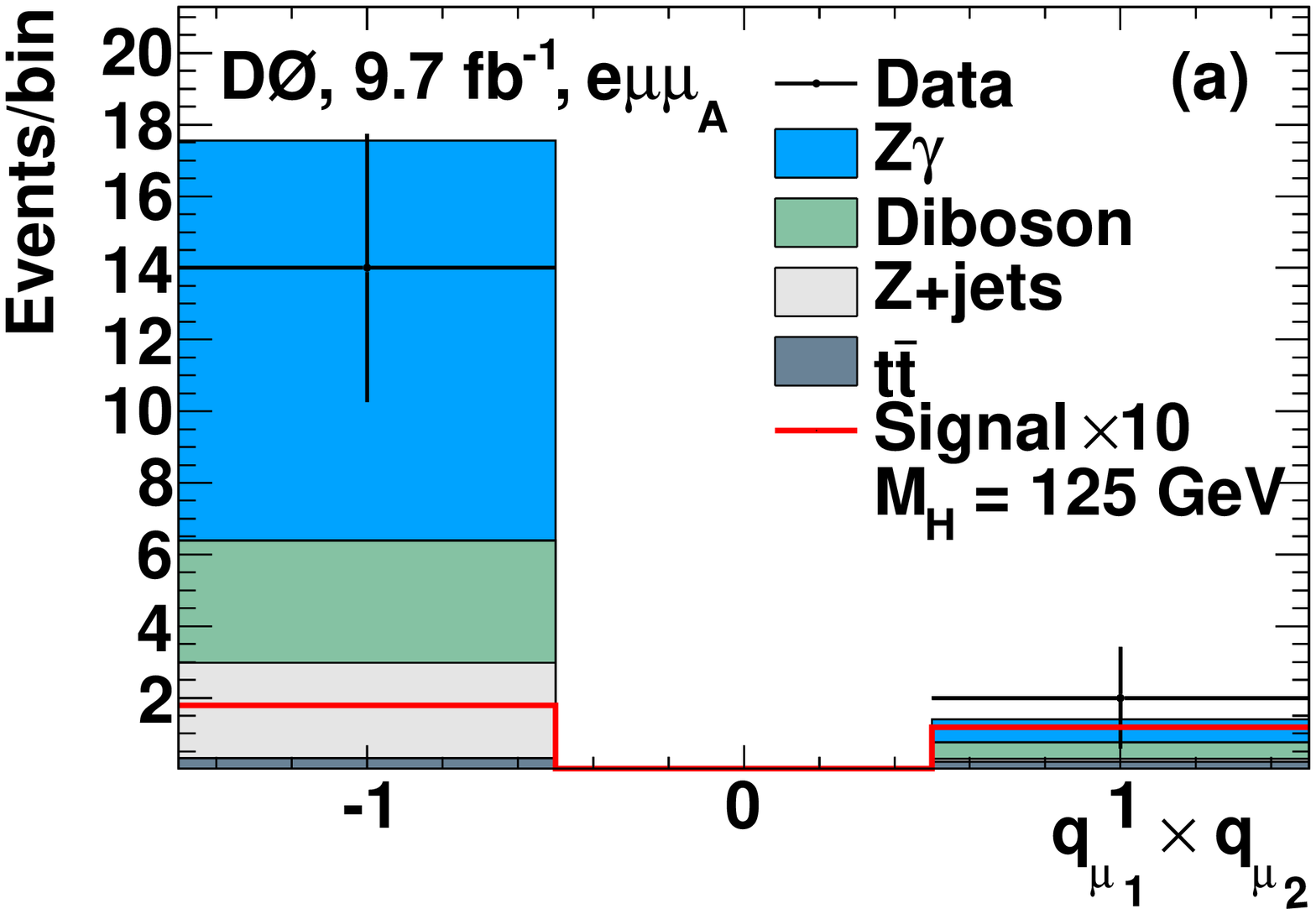}
\includegraphics[scale=0.29]{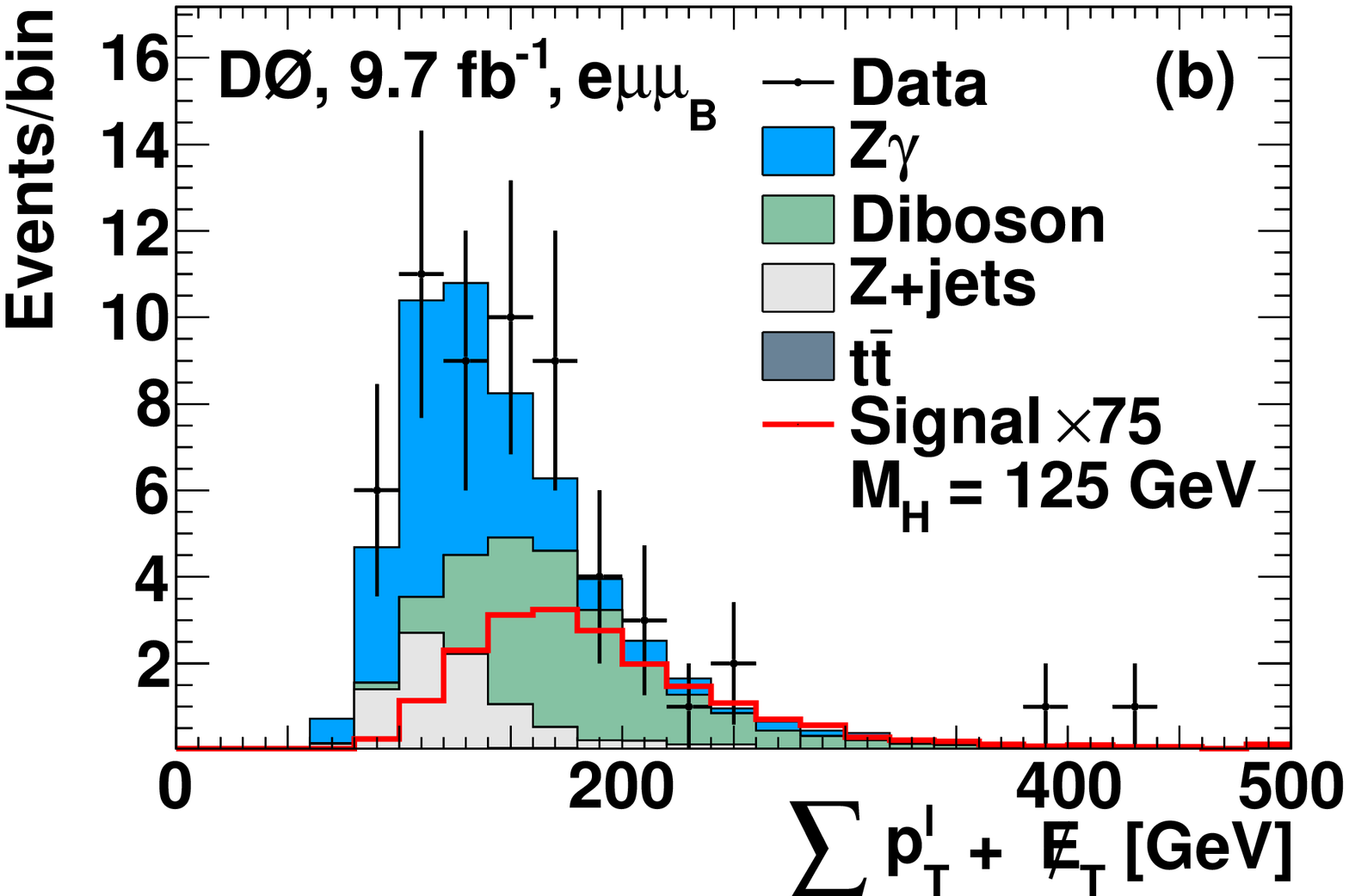}
\includegraphics[scale=0.29]{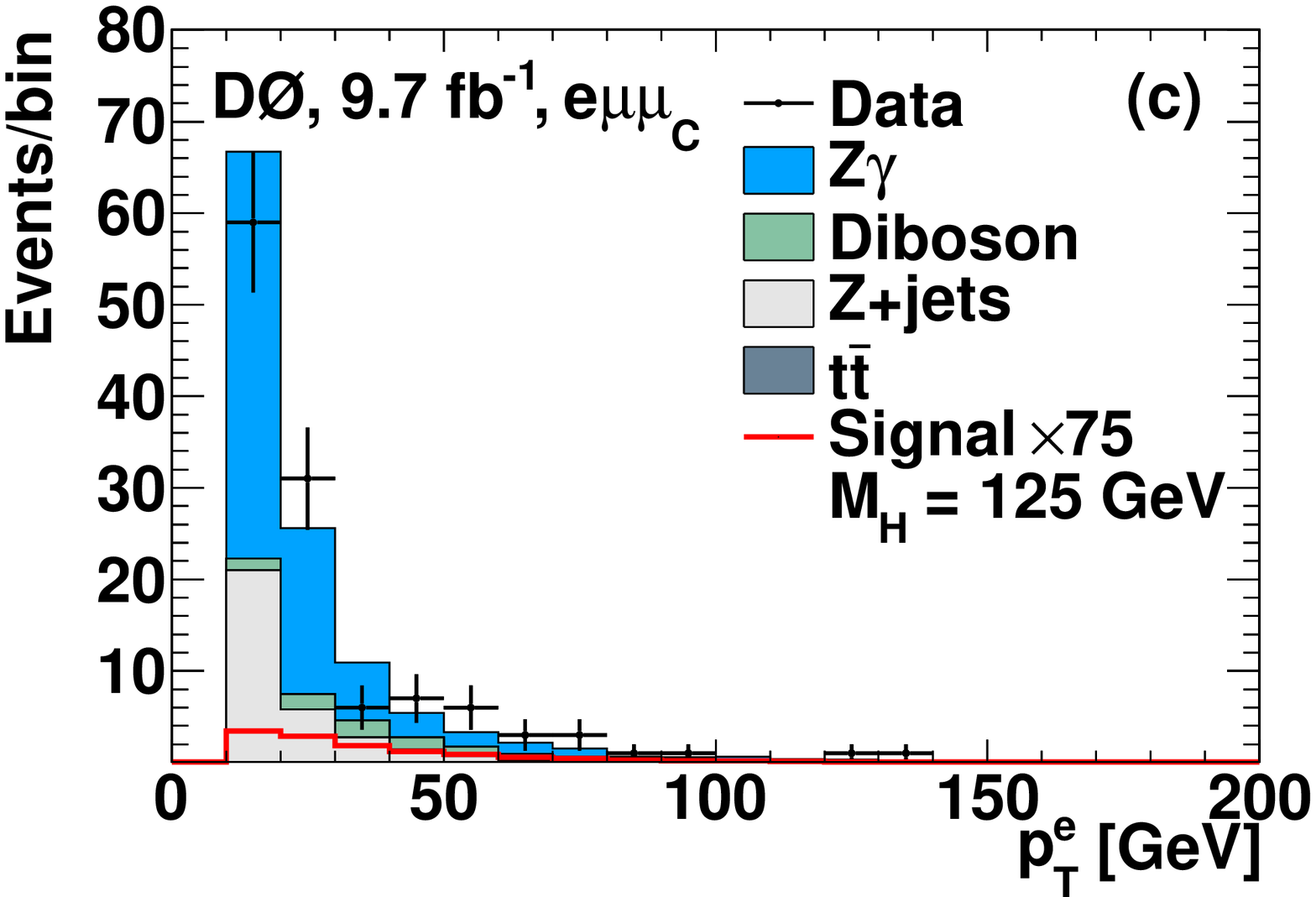}\\
\includegraphics[scale=0.29]{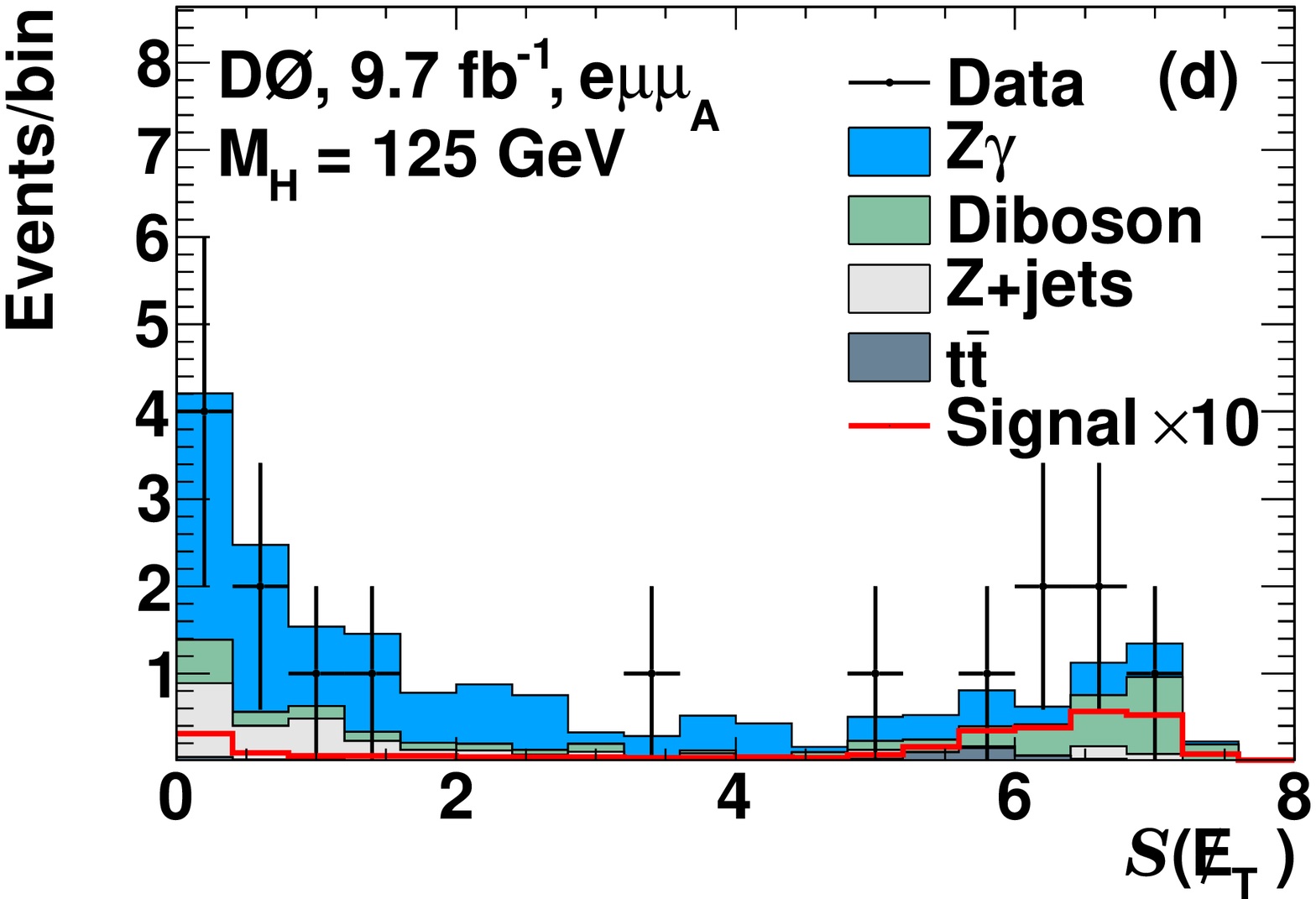}
\includegraphics[scale=0.29]{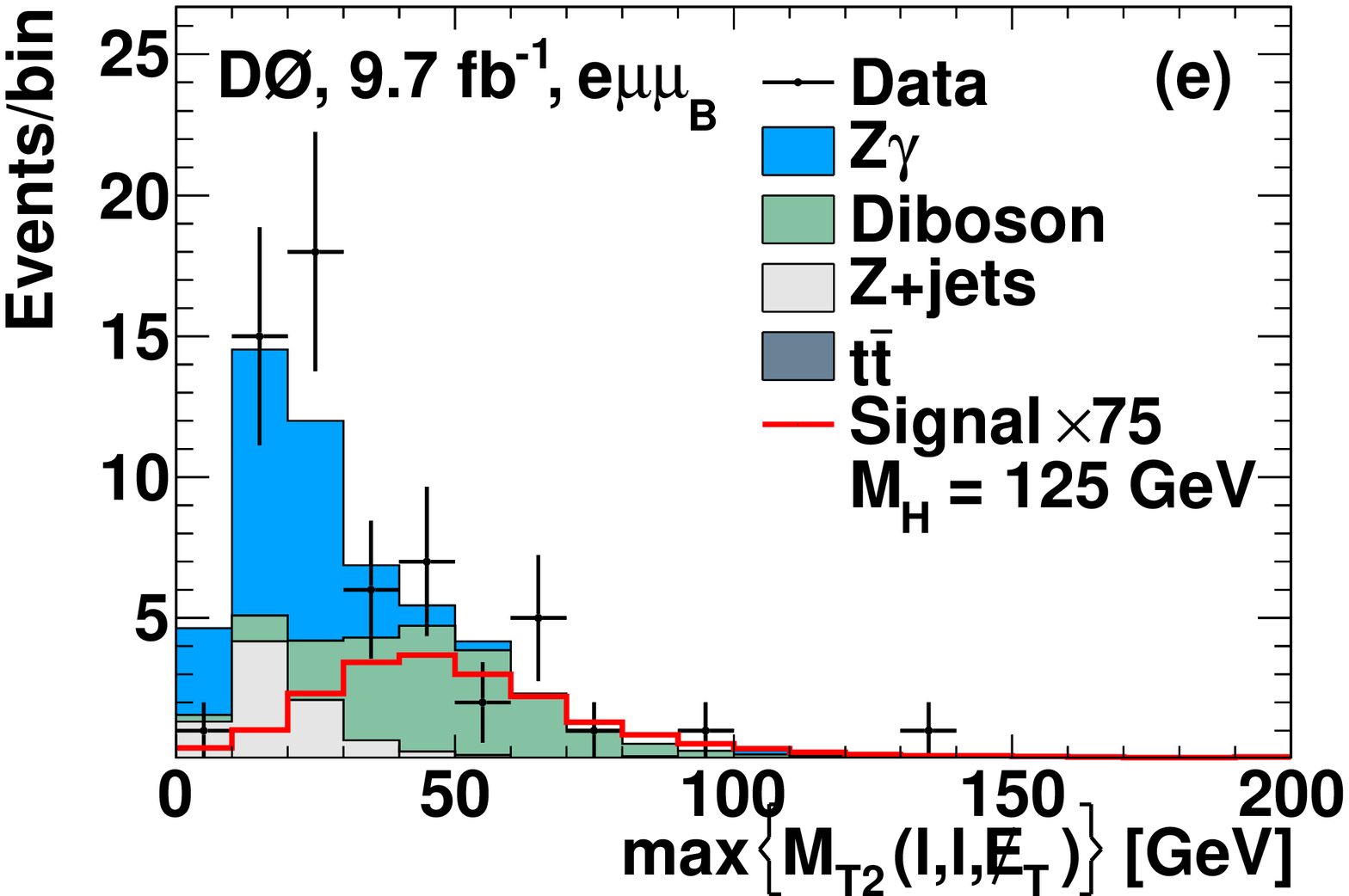}
\includegraphics[scale=0.29]{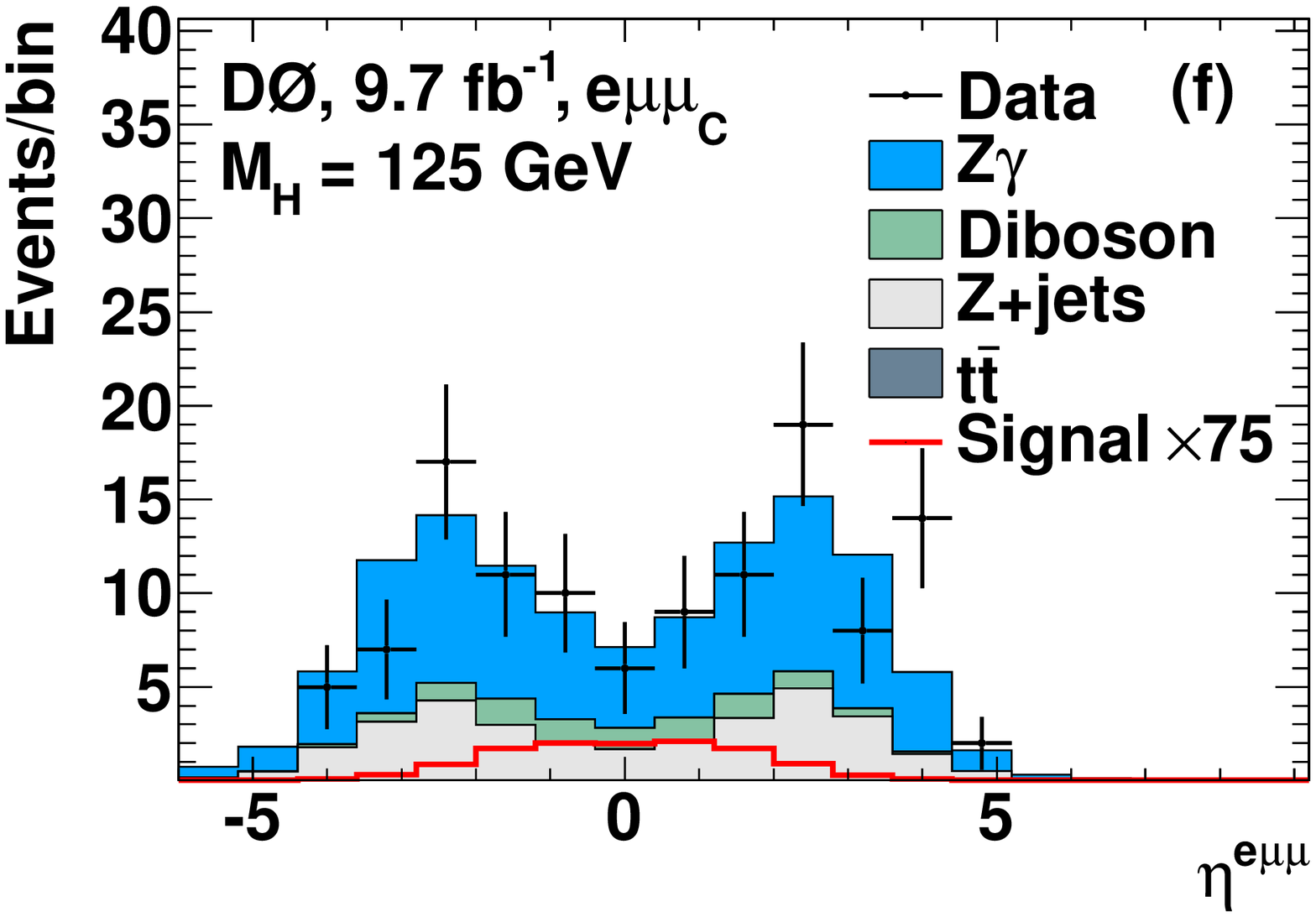}
\caption{\protect\label{fig-kine1} (color online).
Distributions of 
(a) the product of the muon charges, $q_{\mu_1} \times q_{\mu_2}$, 
(b) the scalar sum of transverse momenta and $\MET$, $\sum p_T^{\ell} + \MET$,
(c) the transverse momentum of the electron, $p_T^e$, 
(d) the significance $\METsig$,
(e) the maximal $M_{T2}$ for all lepton pairings, $\max\{M_{T2}(\ell\ell'\MET)\}$,  
and 
(f) the pseudorapidity of the trilepton system, $\eta^{e\mu\mu}$.
The distributions are shown after the event selection.
The $e\mu\mu_{\rm A,B,C}$ samples are shown in the left, middle, and right column,
respectively. 
The data are compared to the sum of the expected background and
to simulations of a Higgs boson signal for $M_H=125$~GeV, multiplied
by factor of 10 for the $e\mu\mu_{\rm A}$ channel and 75 for the other channels.
}
\end{figure*}

\begin{figure*}[hbt]
\includegraphics[scale=0.29]{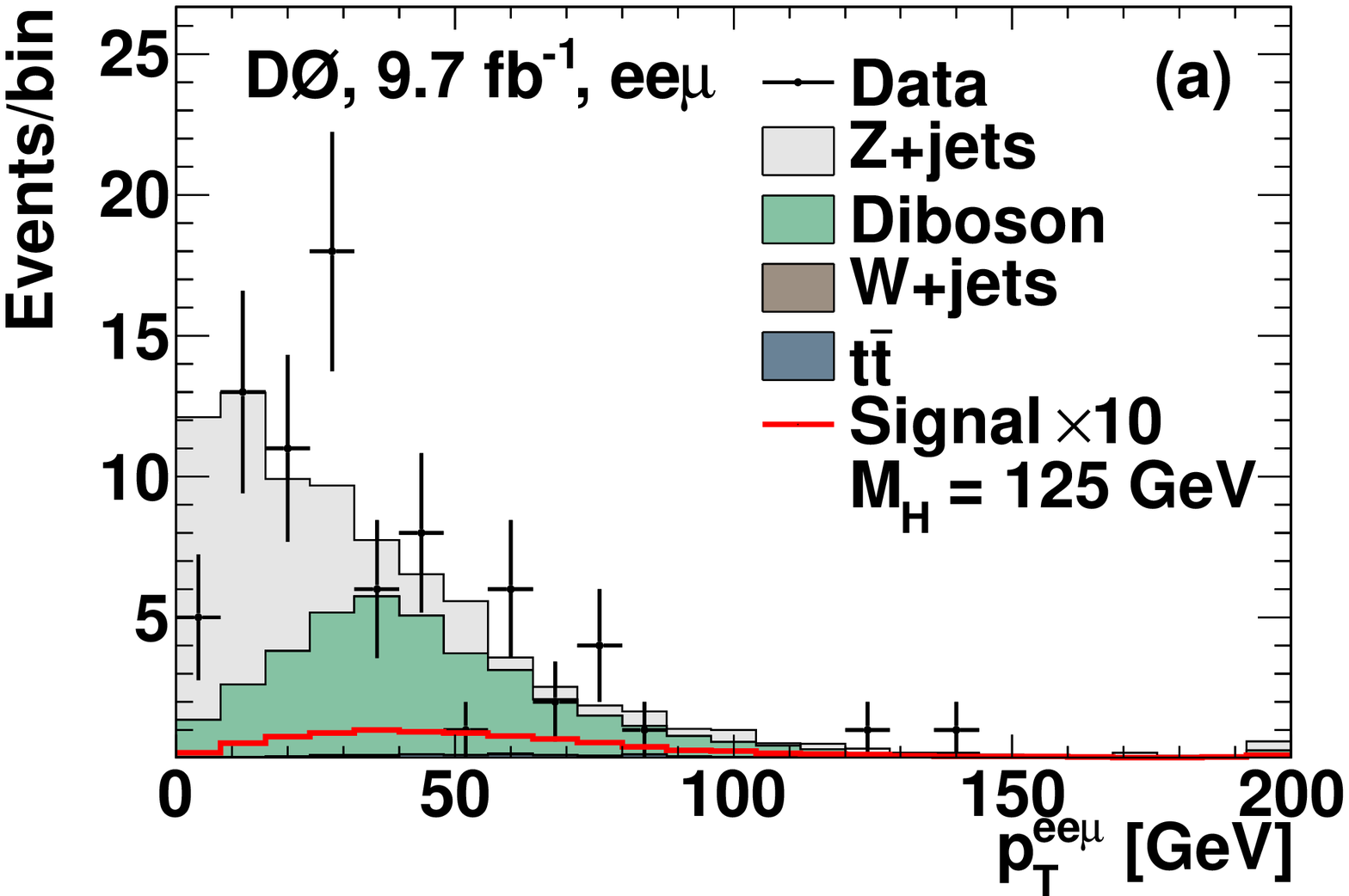}
\includegraphics[scale=0.29]{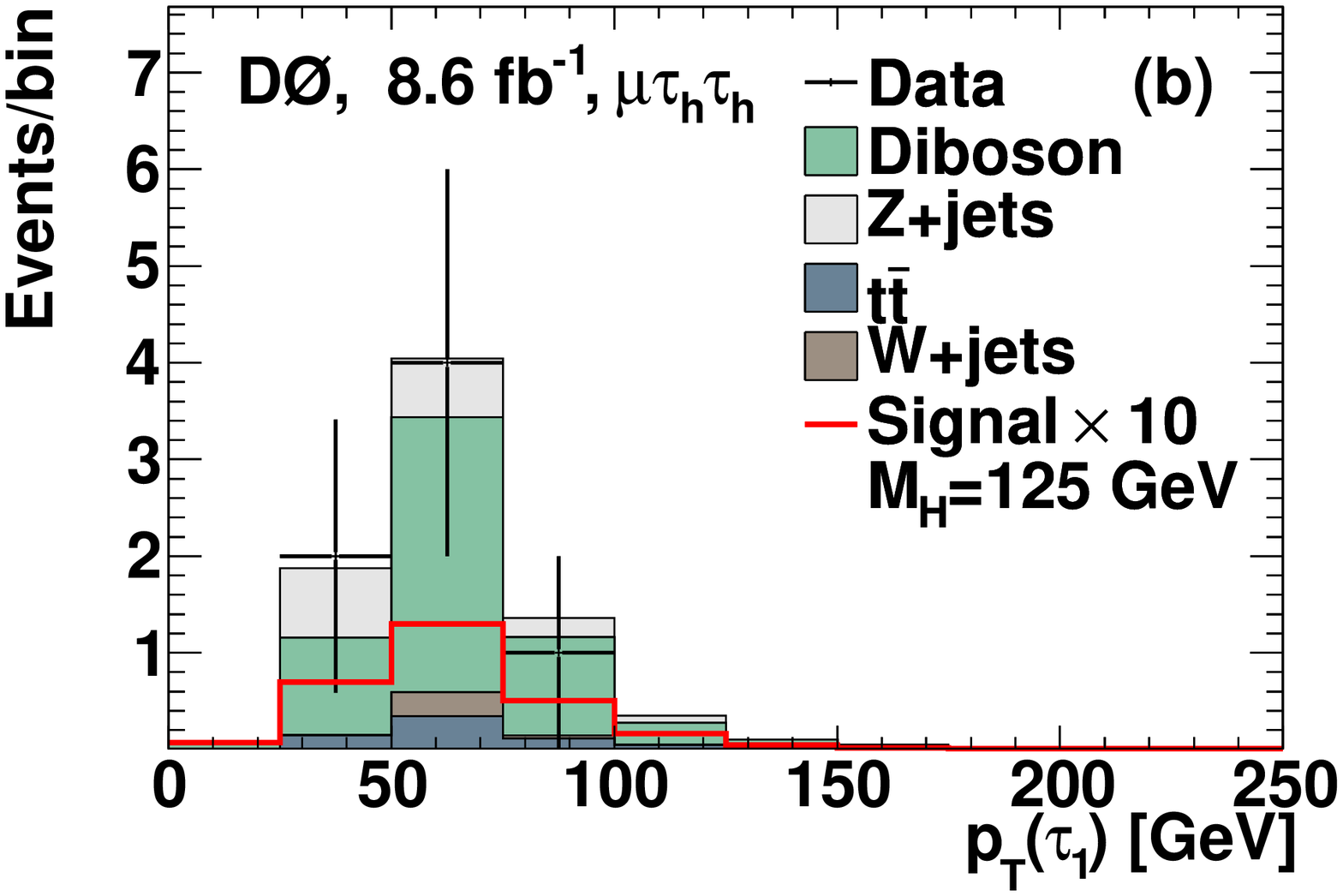}
\includegraphics[scale=0.29]{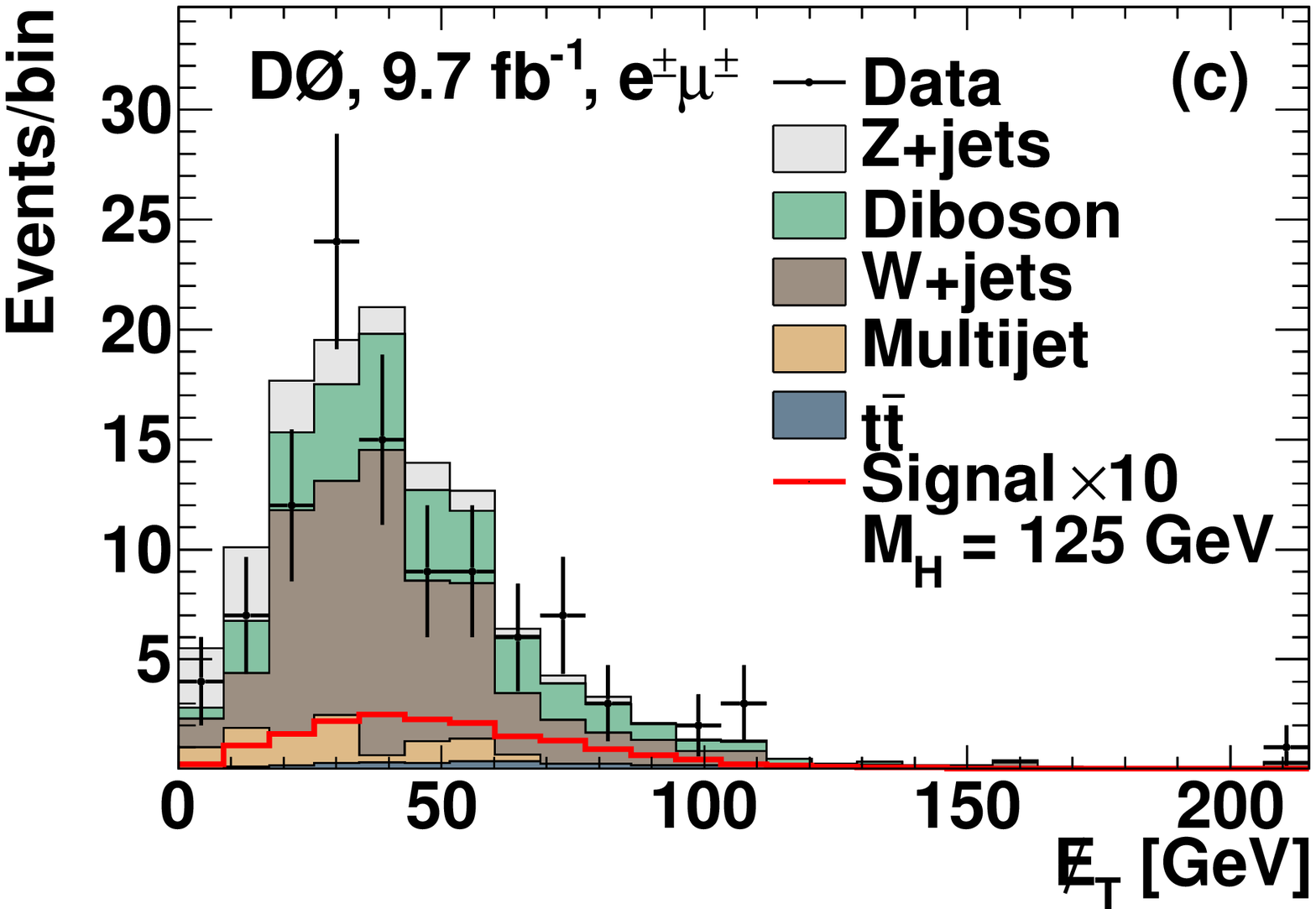}\\
\includegraphics[scale=0.29]{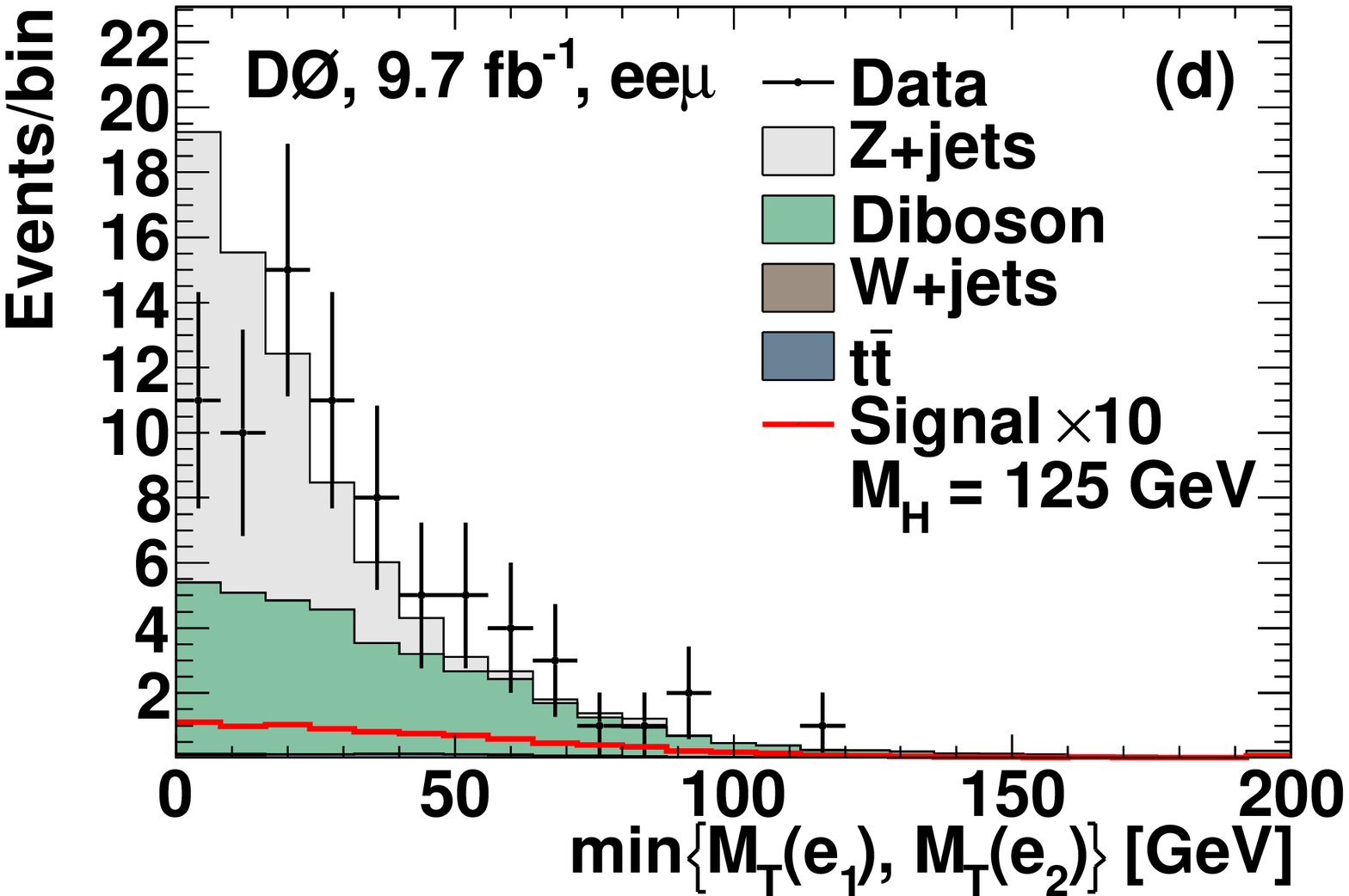}
\includegraphics[scale=0.29]{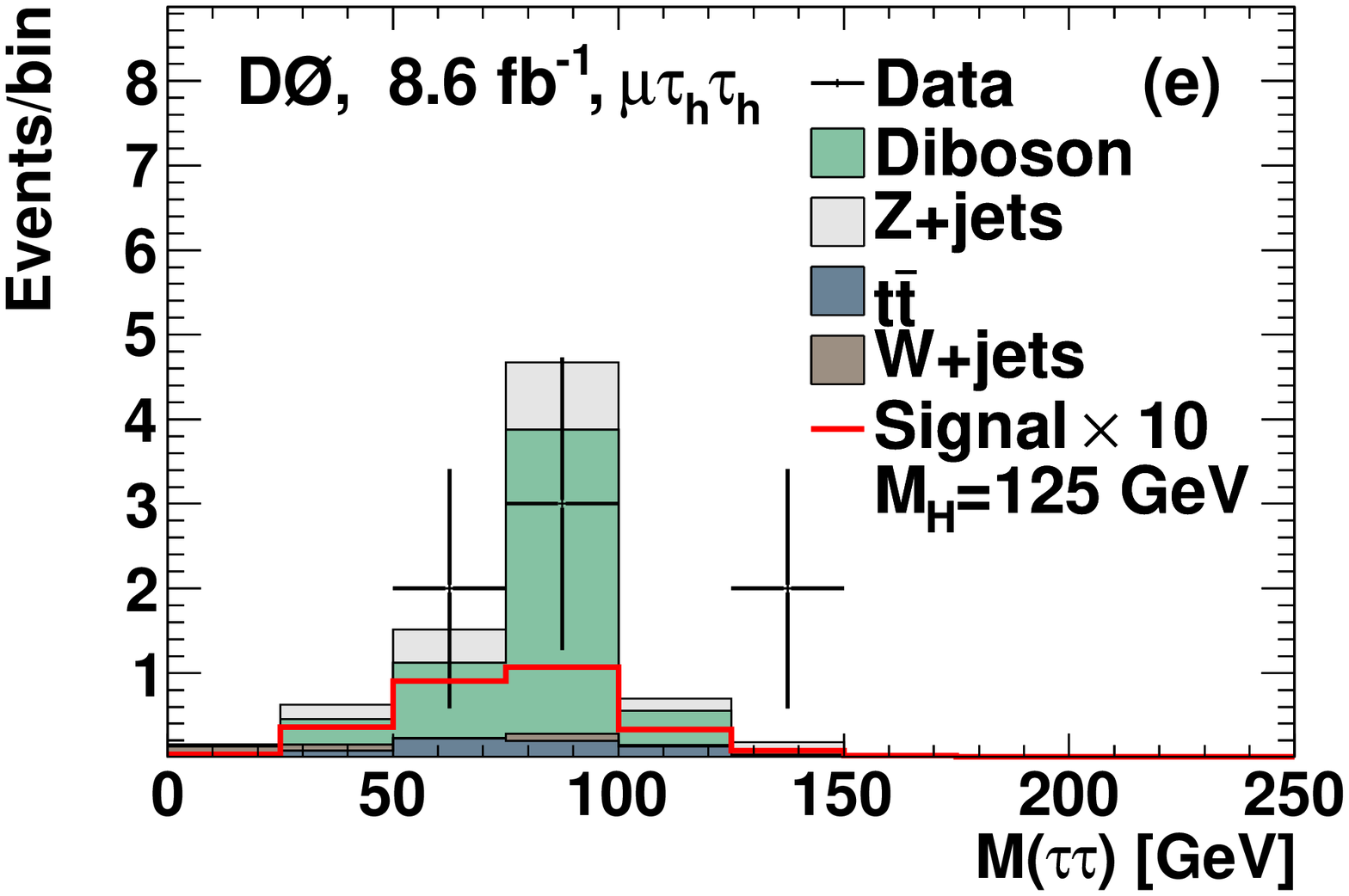}
\includegraphics[scale=0.29]{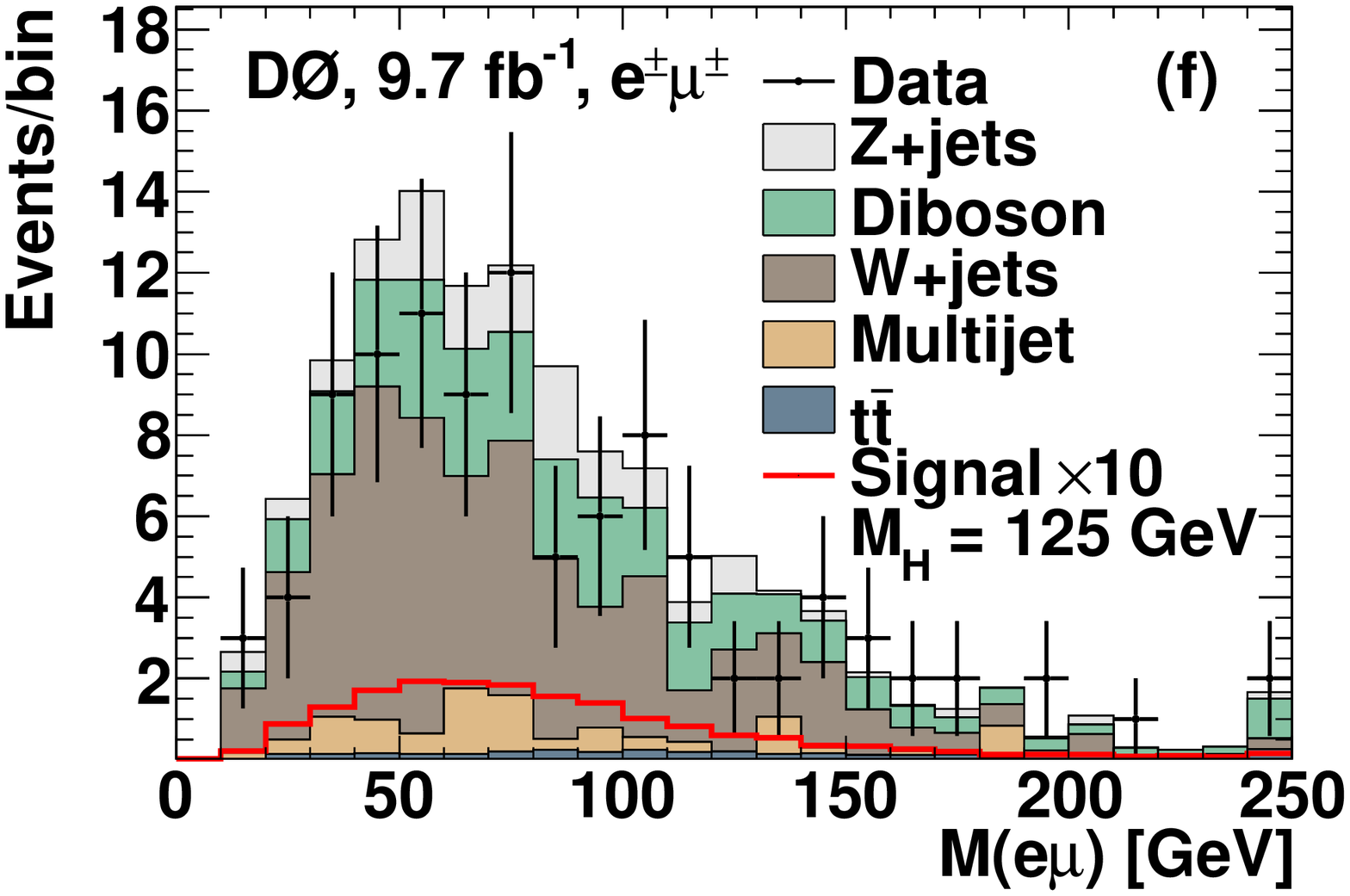}
\caption{\protect\label{fig-kine2} (color online).
Distributions of 
(a) trilepton transverse momentum, $p_T^{ee\mu}$, and (d)
the minimal transverse mass, $\min\{M_T(e_1), M_T(e_2)\}$, for the $ee\mu$ samples,
(b) the transverse momentum of the leading $\tau_h$ candidate,
$p_T^{\tau_1}$, and (e) the di-tau invariant mass, $M(\tau_1,\tau_2)$,
for the $\mu\tau_h\tau_h$ samples, and
(c) missing transverse energy, $\MET$,  and 
(f) the invariant mass of the electron and muon, $M_{e\mu}$,
for the $e^{\pm}\mu^{\pm}$ samples.
The distributions are shown after the event selection. In addition,
the final selection step requiring the output of the first BDT to be $>0.3$ and $\min\{M_T(e),M_T(\mu)\}>7$~GeV
has also been applied to the $e^{\pm}\mu^{\pm}$ sample.
The data are compared to the sum of the expected background and
to simulations of a Higgs boson signal for $M_H=125$~GeV,
multiplied
by a factor of 10.
}
\end{figure*}

Three types of tau lepton decays into hadrons are identified by their signatures.
Type-1 tau candidates consist of a single track and its associated
energy deposit in the calorimeter, without any 
additional separate energy deposits in the EM section. 
This signature corresponds mainly 
to $\tau^{\pm} \rightarrow \pi^{\pm} \nu$ decays and also includes 
leptonic $\tau^{\pm}\to e^{\pm}\nu\nu$ decays.
For type-2 tau candidates, we require a track and its associated
calorimeter energy deposit, plus a separate energy deposit in
the EM calorimeter consistent with a $\pi^0\to\gamma\gamma$ decay, as expected for 
$\tau^{\pm} \rightarrow \pi^{\pm} \pi^{0} \nu$ decays. 
Finally, type-3 tau candidates consist of two or three tracks, 
combined with an energy deposit in the
calorimeter. This corresponds mainly to the decays
$\tau^{\pm} \rightarrow \pi^{\pm} \pi^{\pm} \pi^{\mp} (\pi^{0}) \nu$.
In this analysis, type-3 tau candidates are required to have three tracks 
and an associated net charge of $\pm 1$.
For each tau-type, a neural network is
designed to discriminate $\tau_h$ from jets. The neural network
discriminants are required to be {\it NN}$_{\tau}>0.75$ for types-1 and 2, and
{\it NN}$_{\tau}>0.95$ for type-3~\cite{d0-z-tautau}.
The input variables for these neural networks are based on 
isolation variables for objects
and on the spatial distributions of showers. 

Variables that include information on the imbalance in transverse energy 
($\MET$) caused by neutrinos are used to improve the discrimination
between signal and background.
The $\MET$ is calculated using the transverse energy measured in the 
calorimeter, corrected for the presence of identified muons. 
Two modified $\MET$ variables, 
$\METspecial$ and $\METsig$, are used to reject events 
where the $\MET$ arises from detector effects and not from neutrinos.
In events where the opening angle $\phi$ between the $\MET$ direction
and the nearest lepton or jet is small, the resolution of the $\MET$ measurement
is dominated by the uncertainty on the measured lepton or jet energy. 
Less significance is assigned to this region 
by using  $\METspecial$, defined as $\METspecial=\MET\sin\phi$ if $\phi \le \pi/2$ and
$\METspecial=\MET$ elsewhere.
The significance $\METsig$~\cite{ariel} is defined so that
larger values of  $\METsig$ correspond to $\MET$ measurements that are less likely to be caused by
fluctuations in jet energies.
 
Jet variables are used to discriminate between signal and background
in the $e\mu\mu$ and $e^{\pm}\mu^{\pm}$ channels but not in the event selection.
We identify jets using a midpoint cone algorithm~\cite{jetcone} with a cone size 
of $\mathcal{R} = 0.5$, based  on energy deposits in the calorimeter. 
We require $p_{T}^{\rm jet} > 15$~GeV in the $e\mu\mu$ channel
and $p_{T}^{\rm jet} > 20$~GeV in the $e^{\pm}\mu^{\pm}$ channel.
In both cases the jets must lie within $|\eta^{\rm jet}_{d}| < 2.4$.

\section{Event Selection}

The event selection is designed to maximize sensitivity to a SM Higgs boson signal
in each channel separately. 
The leading muon $(\mu_1)$ in the $e\mu\mu$ and $\mu\tau_h\tau_h$ channels, 
the leading electron ($e_1$) in the $e e \mu$ channel, and the 
electron in the $e^{\pm} \mu^{\pm}$ channel are required to have $p_T>15$~GeV. 
All other selected leptons must have a transverse momentum of $p_T>10$~GeV.
The pseudorapidity of at least one of the selected muons in all channels 
except $ee\mu$ and of both $\tau_h$ candidates in the $\mu\tau_h\tau_h$ channel
must be $|\eta^{\mu}|< 1.6$ and  $|\eta^{\tau}| < 1.5$, respectively. 
The transverse momentum of type-1 and type-2 $\tau_h$ candidates must be $p_T>12.5$~GeV,
and we require $p_T> 15$ GeV for type-3 $\tau_h$ candidates.

The leptons in the events originate from a $\ppbar$ interaction 
vertex, which is required to have a longitudinal coordinate located within $60$~cm 
of the nominal center of the detector. 
The maximum difference between the longitudinal coordinate at the distance 
of closest approach to the beam axis
for all lepton pairings in an event must be less than $3$~cm.

To facilitate combining channels, we ensure that there is no overlap between them.
All events with at least two electrons and at least two muons ($ee\mu\mu$) 
are included in the $e e \mu$ sample and removed from the $e\mu\mu$ sample.  
All events included in the other trilepton final states are removed 
from the $\mu\tau_h\tau_h$ sample.  
We also
reject events with an additional electron or muon in the $e^{\pm} \mu^{\pm}$ channel.

We construct a variable $M(\ell\ell\ell\MET)$ that is the invariant mass
of the three leptons and the $\MET$, where the $\MET$ vector is assumed to have 
a longitudinal momentum component equal to zero. 
We require $M(\ell\ell\ell\MET) >100$~GeV for the $ee \mu$ and $e\mu\mu$  
final states to reject $Z$+jets background. 
To remove $Z \to \mu \mu$ events with final state radiation, 
we require $e\mu\mu$ events in the range
$75<M(e\mu\mu)<105$~GeV to have $\MET> 20$~GeV.  
In the $\mu\tau_h\tau_h$ final state, 
we require $\MET>15$~GeV and a transverse mass $M_T(\mu)>20$~GeV. 
The transverse mass 
$$M_T(\ell)= \sqrt{2  p_T^{\ell} \MET (1 - \cos\phi))}$$
is calculated using the azimuthal angle $\phi$ between the charged lepton ($\ell=e,\mu$)
and the direction of the $\MET$. 

Muons and electrons must be separated from jets by 
$\mathcal{R} > 0.1$ in the $e\mu\mu$, $ee\mu$, and $e^{\pm}\mu^{\pm}$ 
channels.
All selected leptons are required to be separated by $\mathcal{R} > 0.3$ from each other.  
This is increased to $\mathcal{R} > 0.5$ for the pairings of $\tau_h$ candidates and 
for the pairing between $\tau_h$ candidates and muons. 
The sum of the charges in the $\mu\tau_h\tau_h$ final state must be $\pm 1$. The electron
and muon are required to have the same charge in the $e^{\pm}\mu^{\pm}$ final state.
No lepton charge requirements are applied for the $ee \mu$ or $e\mu\mu$ final states
to maximise the sensitivity to signal.

We divide the $e\mu\mu$ channel into three samples with different signal
and background composition to increase sensitivity to a Higgs boson signal.  
The $e\mu\mu_{\rm A}$ sample contains events where the 
dimuon mass is outside the range $60<M(\mu\mu)<130$~GeV and all events
with like-charge muons. 
The second sample ($e\mu\mu_{\rm B}$) contains 
events with oppositely charged muons, $60<M(\mu\mu)<130$~GeV, and $\METsig > 2$.  
The $e\mu\mu_{\rm C}$ sample consists of all remaining events with 
$60<M(\mu\mu)<130$~GeV  and $\METsig \leq 2$. 

The number of events in data, the expected background and signal, 
after all selection criteria described in this Section have been applied,
are given in Table~\ref{tab:cutflow}.

\section{Instrumental Backgrounds \label{instrumbkg}}

Instrumental backgrounds are caused by leptons produced inside
jets, low-multiplicity jets that are reconstructed as $\tau_h$ candidates, 
photons or jets misidentified as electrons, and by
opposite-charge $e\mu$ pairs where one of the charges is incorrectly measured.  

The $W$+jets background in the $e\mu\mu$ and $ee\mu$ samples is expected to be
small, and its contributions is therefore described only by the simulation using
the theoretical cross section.
Since the $W$+jets background is expected to contribute more 
for the $\mu\tau_h\tau_h$ and $e^{\pm} \mu^{\pm}$ final states,
their normalisation is obtained using data and then applied to
the simulated kinematic distributions of the $W$+jets events.

To model the $W$+jets background for $\mu\tau_h\tau_h$ 
final states,  we select a data sample enriched in $W$+jets events. 
We require that events pass all selection criteria,
except the requirements on the {\it NN}$_{\tau}$ outputs. In addition, we require
$M_T(\mu) > 40$~GeV, and $p_T^{\tau_1} > 20$~GeV. 
The normalization factors that are applied to the simulation are determined from the ratio of the event
yields in the $W$+jets enriched region for data and simulation. They
are determined separately for each type of $\tau_h$ candidate, and for same-charge
and opposite-charge $\tau_h$ pairs.

To normalize the simulated $W$+jets background 
in the $e^{\pm} \mu^{\pm}$ final state, we select a data sample
enriched in $W$+jets events by requiring
$\MET > 20$~GeV, an inverted electron likelihood criterion $\mathcal{L}_e < 0.7$, 
and $\min\{M_T(e),M_T(\mu)\} > 20$~GeV.
This data sample is used to derive separate normalization factors 
for jet multiplicities $N_{\rm jet}=0,1$, and $\geq 2$ that are applied to the 
simulated $W$+jets background samples.

Multijet background is negligible for the $e\mu\mu$, $e e \mu$, and
$\mu\tau_h\tau_h$ final states.  The multijet background
in the $e^{\pm}\mu^{\pm}$ channel is determined using a sample
with relaxed lepton identification requirements. The events in this sample
contain one electron with lower requirements on the likelihoods and neural networks
compared to the standard electron identification used in the event selection, and
one muon without isolation requirements applied.

To obtain the correct normalization of this sample, we calculate separate fake rates 
for electrons and muons, given by the ratio of the number of events with a standard
lepton to the number of events with a fake lepton. Fake leptons
are defined by applying the relaxed lepton identification requirements 
but removing standard leptons.
This fake rate is calculated separately for electrons and muons as a function of their 
transverse momentum in a sample
with $\MET<15$~GeV. This selection ensures that both samples used 
to calculate the fake rates are enriched in multijet events.
The shape and normalization of the multijet contribution is then
obtained by applying the product of the fake rates for electrons and muons
to the multijet data sample selected with the relaxed lepton identification requirements.

\begin{table*}
\caption{\protect\label{tab:BDTinput} 
Set of variables used in training of the BDT for each final state.
The charges of the leptons are $q_{\ell}$ ($\ell=e,\mu$). 
The angle $\phi(\ell_1,\ell_2)$ is taken between the two leptons, and the
angle  $\phi(\ell\ell,\ell')$ between the dilepton system ($\ell\ell$) and the lepton
with the different flavour ($\ell'$).
The variables $\cal{R(\ell,\ell)}$ and $M_{T2}(\ell\ell'\MET)$ are calculated for all lepton pairings.
The  pairing with the smallest and largest values are denoted by 
 $\min\{\, \}$ and $\max\{\, \}$, respectively, and mid$\{\,\}$ corresponds to the third pairing.
 The mass, transverse momentum, and pseudorapidity of the trilepton system are denoted by 
 $M(\ell\ell\ell)$, $p_T^{\ell\ell\ell}$,
 and $\eta^{\ell\ell\ell}$, respectively, and $\sum p_T^{\ell}$ is the scalar
sum of the transverse momenta of the three leptons.  
The variable $f_{cp}$ is the fraction of charged particle tracks 
associated with the jet that point back to
the same vertex as the leptons. If no jets are present, the jet variables are set to zero.
All variables given for the $\mu\tau_h\tau_h$ channel are used in the first BDT, whereas only
$p_T^{\tau_1}$, $M(\tau\tau)$, $M_T(\mu)$, and $\MET$ are used in
the second BDT.
All variables given for the $e^{\pm}\mu^{\pm}$ channel are used in both
BDTs, except  $p_T^{\rm jet}$ and $\phi(\mbox{jet},\MET)$, which are used in 
the second BDT only.}\begin{center}
\begin{tabular}{c|c|c|c|c|c|c|c} 
\hline
\hline
 $ee\mu$ & $e\mu\mu_{\rm A}$ & $e\mu\mu_{\rm B}$ & $e\mu\mu_{\rm C}$ & \multicolumn{2}{c|}{$\mu\tau_h\tau_h$} & \multicolumn{2}{c}{$e^{\pm}\mu^{\pm}$}  \\
 & & & & \multicolumn{1}{c}{BDT1} & BDT2 &  \multicolumn{1}{c}{BDT1} & BDT2 \\
\hline
&      &                           &       &  &  &  \multicolumn{2}{c}{$\mathcal{L}_e$}  \\
$p_T^{e_1}$, $p_T^{e_2}$, $p_T^\mu$ &$p_{T}^{e} $  & $p_{T}^{e} $  
&$p_T^{\mu_1}$,$p_{T}^{e}$  & $p_T^{\tau_1}$, $p_T^{\tau_2}$, $p_{T}^{\mu} $ & $p_T^{\tau_1}$
 &\multicolumn{2}{c}{$p_T^{e}$,$p_T^{\mu}$}   \\
$p_T^{ee}$,$p_T^{ee\mu}$ & $p_T^{\mu\mu}$ &   $p_T^{\mu\mu}$ &   $p_T^{e\mu\mu}$ & $p_T^{\tau \tau}$ & & \multicolumn{2}{c}{} \\
& $\eta^{e\mu\mu}$ &   $\eta^{e\mu\mu}$ & $\eta^{e\mu\mu}$ & & & \multicolumn{2}{c}{} \\
& $q_{\mu_1} \times q_{\mu_2}$ & & & & & 
\multicolumn{2}{c}{$q_e\times\eta^e$, $q_{\mu}\times\eta^{\mu}$}
  \\
$\phi (e_1,e_2)$       & $\phi (\mu_1, \mu_2)$        & $\phi (\mu_1, \mu_2)$ 
&$ \phi (\mu_1, \mu_2)$    & &   & \multicolumn{2}{c}{}\\
$\phi (ee,\mu)$  & $\phi (\mu\mu,e)$ &    $\phi (\mu\mu,e)$
& $\phi (\mu \mu,e)$ &  & &  \multicolumn{2}{c}{} \\        
$\min\{\mathcal{R}(\ell,\ell')\}$ & $\min\{\mathcal{R}(\ell,\ell')\}$ & $\min\{\mathcal{R}(\ell,\ell')\}$ 
& $\min\{\mathcal{R}(\ell,\ell')\}$  & & & \multicolumn{2}{c}{$\mathcal{R}(e,\mu)$} \\
mid$\{\mathcal{R}(\ell,\ell')\}$ & & & mid$\{\mathcal{R}(\ell,\ell')\}$ &  & &
\multicolumn{2}{c}{} \\
    & &  & $\max\{\mathcal{R}(\ell,\ell')\}$  & & & \multicolumn{2}{c}{} \\
$M(ee)$   & $M(\mu\mu)$ &  $M(\mu \mu)$ & $M(\mu \mu)$ & $M(\tau \tau)$, $M(\mu \tau_1)$ & $M(\tau \tau)$ & \multicolumn{2}{c}{$M(e\mu)$}  \\
$M(ee\mu \MET)$           & $M(e\mu\mu)$   & &   $M(e\mu\mu)$  & $M(\tau\tau \mu)$ & & 
\multicolumn{2}{c}{} \\ 
$\min\{M_T(e_i)\}$ & $\min\{M_T(\mu_i)\}$ & &   & $M_T(\mu)$ & $M_T(\mu)$ 
&\multicolumn{2}{c}{$\min\{M_T(e),M_T(\mu)\}$}\\
&   $\max\{M_{T2}(\ell\ell'\MET)\}$        &$\max\{M_{T2}(\ell\ell'\MET)\}$  & &  & & \multicolumn{2}{c}{}  \\
&           mid$\{M_{T2}(\ell\ell'\MET)\}$                              &      &     & & &
 \multicolumn{2}{c}{} \\
$\METspecial$, $\METsig$ & $\METspecial$, $\METsig$ & $\METspecial$, $\METsig$ & & $\MET$ & $\MET$ & 
\multicolumn{2}{c}{$\MET$}  \\
&          &$M_T(e,\MET)$                                               & &         & &   \multicolumn{2}{c}{}  \\
&     $\sum p_T^{\ell} + \MET$   &        $\sum p_T^{\ell} + \MET$          &$\sum p_T^{\ell} + \MET$ & &    &  \multicolumn{2}{c}{}   \\
&            &$\min\{\phi(\mu_i,\METnoSpace)\}$&              &  & & 
\multicolumn{2}{c}{$\max\{(\phi(\ell,\MET)\}$}\\   
&  & $N_{\rm jet}$ & $N_{\rm jet}$ & & & \multicolumn{2}{c}{$N_{\rm jet}$}\\ 
\cline{7-8}   
& $p_T^{\rm jet}$ &  $p_T^{\rm jet}$ & $p_T^{\rm jet}$ & & & & $p_T^{\rm jet}$ \\
& & & $\eta^{\rm jet}$ & & & & \\
& & $\max\{\phi(\mbox{jet}_i,\MET)\}$ &    & & &  $\phantom{\phi(\mbox{jet},\MET)}$ &  $\phi(\mbox{jet},\MET)$ \\
& & &  $f_{cp}$ & & & & \\
\hline
\end{tabular}        
\end{center}     
\end{table*}

Photons reconstructed as electrons contribute to the background in the  
$e\mu\mu$ final state. To estimate the contribution from the $Z\gamma$ background,
we select events with two muons, and with a photon or electron (labeled $e/\gamma$)
identified using identical criteria, apart from the track matching requirement, which is reversed for the photon.
In addition, we require $\MET<20$~GeV. The invariant mass of
the two muons and the $e/\gamma$ has to be in the range between $75$ and $105$~GeV.
This sample is enriched in events with final state radiation from 
$Z/\gamma^* \to \mu^+ \mu^-$ decays.
 A normalization factor is calculated
as the ratio of the number of events with electrons to photons in this sample. 
To estimate the $Z\gamma$ background, this normalization factor is
applied to a $\gamma\mu\mu$ data sample selected 
in the same way as the $e\mu\mu$ sample, except for the reversed track matching
requirement used to define the photon.

\section{Multivariate Discriminants}

Boosted decision trees (BDTs), as implemented in the {\sc tmva} package~\cite{bib-tmva},
are used to discriminate between signal and background.
The BDTs are trained for each mass point separately in the range $100\le M_H\le 200$~GeV, in
steps of $5$~GeV.  For each background or signal process, the event samples are split into
subsamples for training the BDTs.
The BDTs are then applied to the subsamples not used in the training to derive 
limits on the Higgs boson production cross section.
The only exception is the $Z\gamma$ background in the $e\mu\mu$ final state, where
both simulation and data are used.
We train the BDT with a $Z\gamma$ MC sample and then apply the BDT to
the kinematic distributions estimated using the $Z\gamma$ data sample to 
obtain the BDT distribution used in the limit setting. 
This procedure reduces fluctuations in the training 
caused by the small number of data events.

The BDTs exploit kinematic differences between Higgs boson production
for a given $M_H$ and background. 
The variables used as inputs to the BDTs are given in Table~\ref{tab:BDTinput}.  
They are based on the transverse momenta of the leptons and jets, the $\MET$, angular
variables, charge correlations between leptons,
and on the invariant masses of the pairings of leptons, jets, and the $\MET$.
Jet variables are used to calculate
the discriminants in the $e\mu\mu$ and $e^{\pm}\mu^{\pm}$ channels only.
The variable $M_{T2}$ is an extension of the transverse mass 
$M_T$ to final states with two visible and two invisible particles~\cite{lester,cheng}. 
Some of the variables use constraints given by the $Z$ boson mass. 
On average, the opening angle between leptons in $H \to WW$ decays are smaller than 
for background and their direction is opposite to the direction of the 
$\MET$ because of spin correlations in the decay of a scalar Higgs boson.
Other variables, such as the likelihood  $\mathcal{L}_{e}$, reject events with misidentified leptons.
Distributions of some of the input variables used for BDT training are shown in 
Figs.~\ref{fig-kine1} and~\ref{fig-kine2}.

\begin{table}
\caption{\protect\label{tab:cutflow2}
Numbers of events in data, predicted background, and expected 
signal for $M_H=125$~GeV for the $e^{\pm}\mu^{\pm}$ channel.
The numbers are given after the initial event selection and after
the final selection, which also requires the first BDT output to be $>0.3$ and 
$\min\{M_T(e),M_T(\mu)\}>7$~GeV. 
The numbers of events are given
 with their total (statistical and systematic) uncertainties.}
\renewcommand{\arraystretch}{1.0}
 \begin{center}
\begin{tabular}{l|c|c} 
\hline
\hline
& Initial Selection & Final Selection \\
\hline
Signal  &&\\
{\sl WH}         & $1.93$  &  $1.51$\\
{\sl ZH}          & $0.32$  &  $0.23$ \\
$gg\to H\to ZZ$     & $<0.01$ & $<0.01$ \\
\hline
Signal Sum  & $2.25$   & $1.74$\\
\hline
Background &&\\
$Z\to e^+e^-$               & $15.9 \pm 2.4\pz$ & $2.7  \pm 0.4 $\\
$Z\to \mu^+\mu^-$       & $58.5 \pm 15.2$  & $10.6 \pm 2.8\pz$\\
$Z\to \tau^+\tau^-$       & $22.0 \pm 6.8\pz$  & $1.8  \pm 0.6$\\
$Z\gamma$                & $<0.1$ & $<0.1$        \\
Diboson          & $36.2 \pm 3.6\pz$ & $31.6 \pm 3.2\pz$\\
$\ttbar$           & $4.1 \pm 2.1$  & $3.4  \pm 1.7$\\
$W$+jets         & $238.3 \pm 19.0\pz$  & $62.4 \pm 5.0\pz$ \\
Multijet             & $434.5 \pm 87.0\pz$ & $9.1  \pm 1.8$\\   
\hline
Background &&\\
Sum           &  $809 \pm 93$ & $122 \pm 7\pz$\\ \hline
Data                     & $822$      & $102$ \\ 
\hline \hline
\end{tabular}
\end{center}
 \renewcommand{\arraystretch}{1.0}
 \end{table}
 
\begin{figure*}[htb]
\includegraphics[scale=0.295]{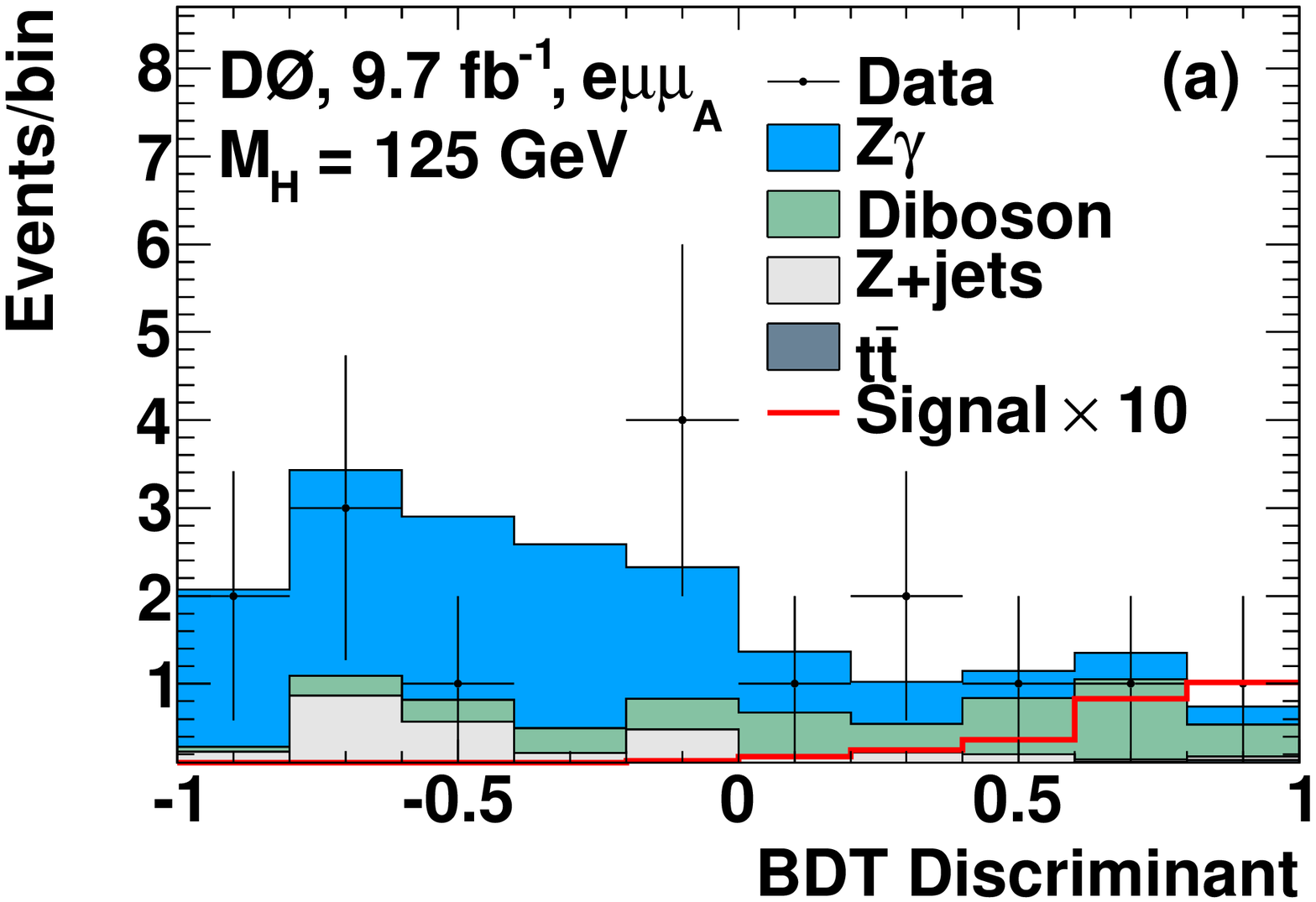}
\includegraphics[scale=0.295]{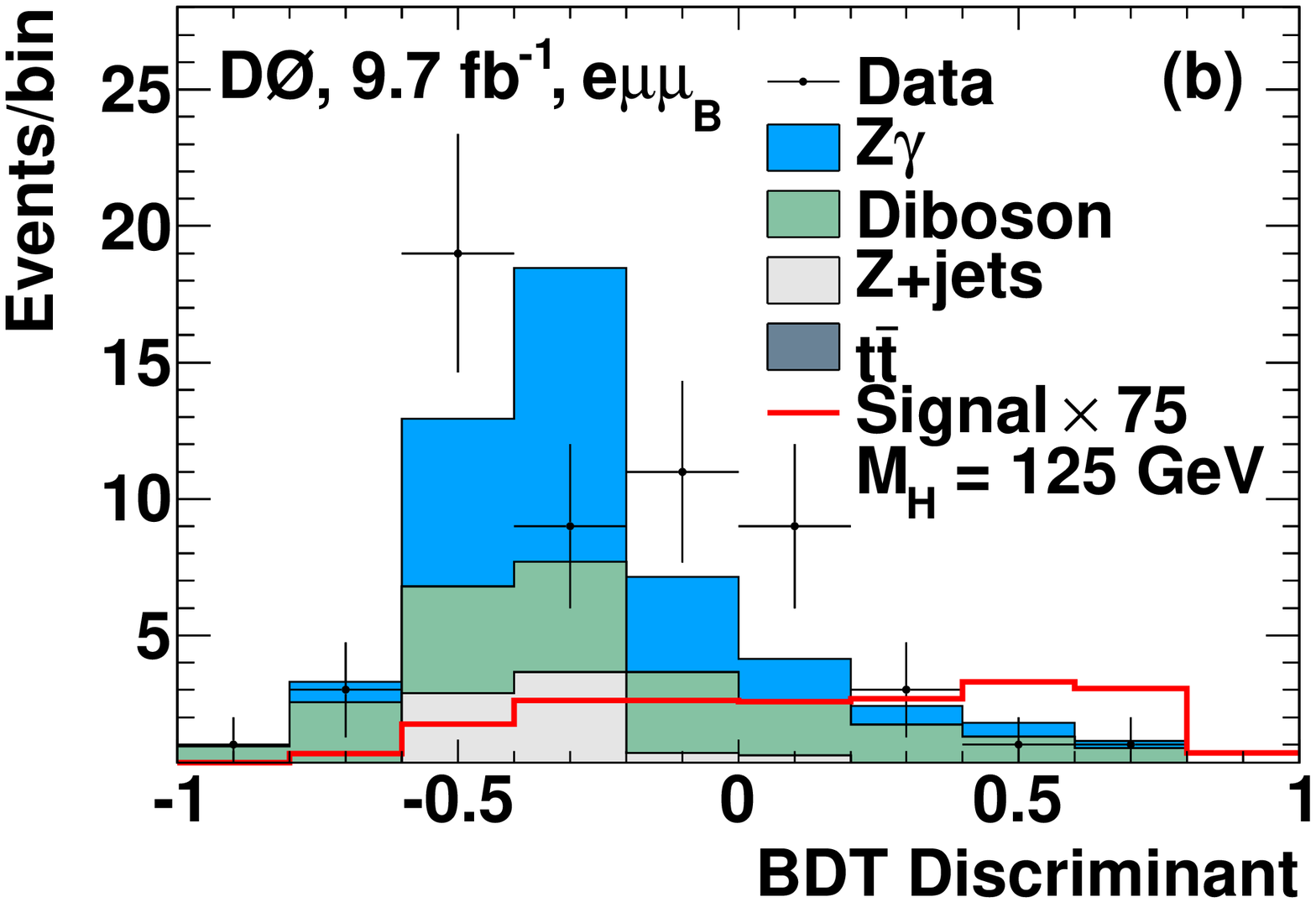}
\includegraphics[scale=0.295]{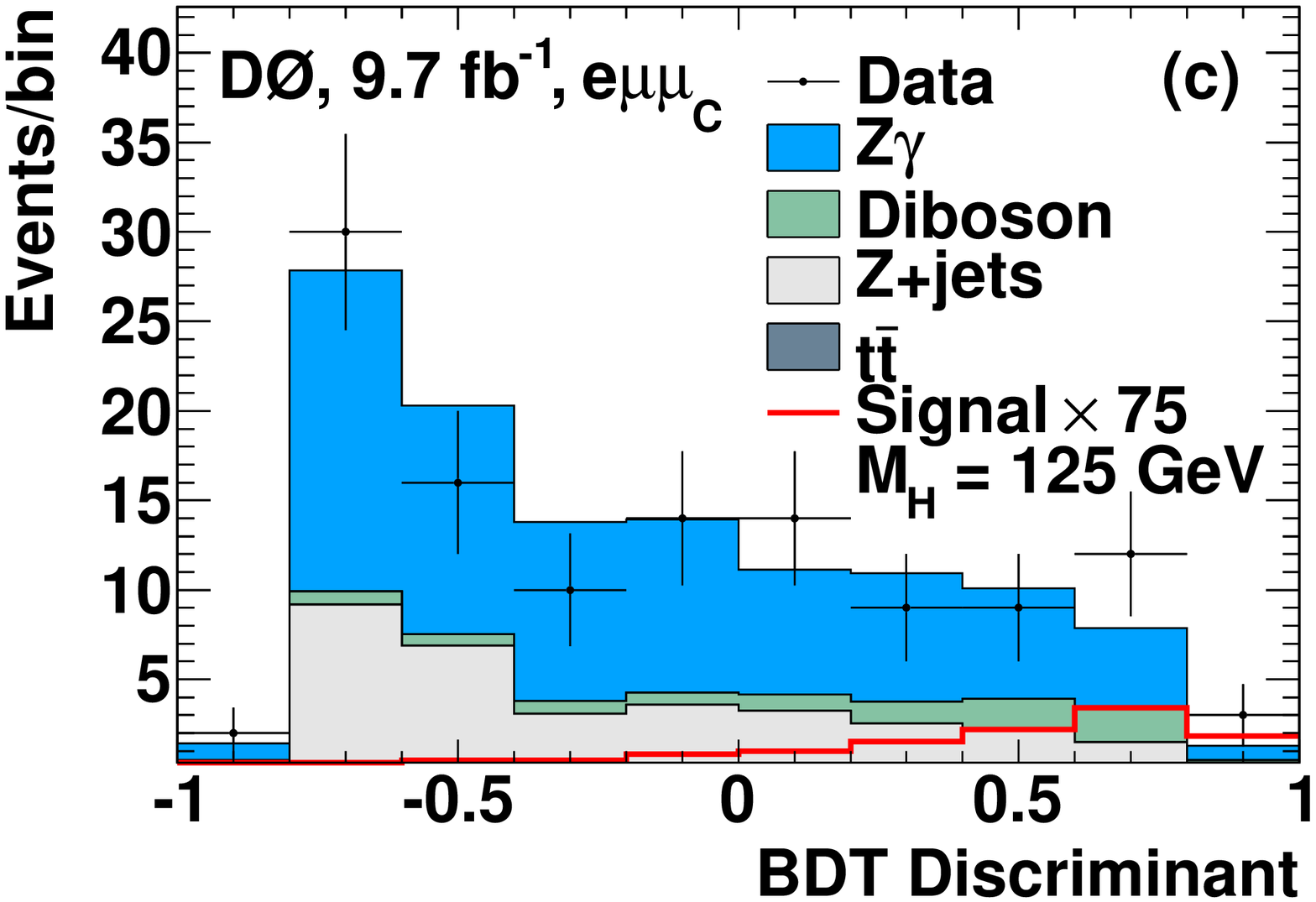}\\
\includegraphics[scale=0.288]{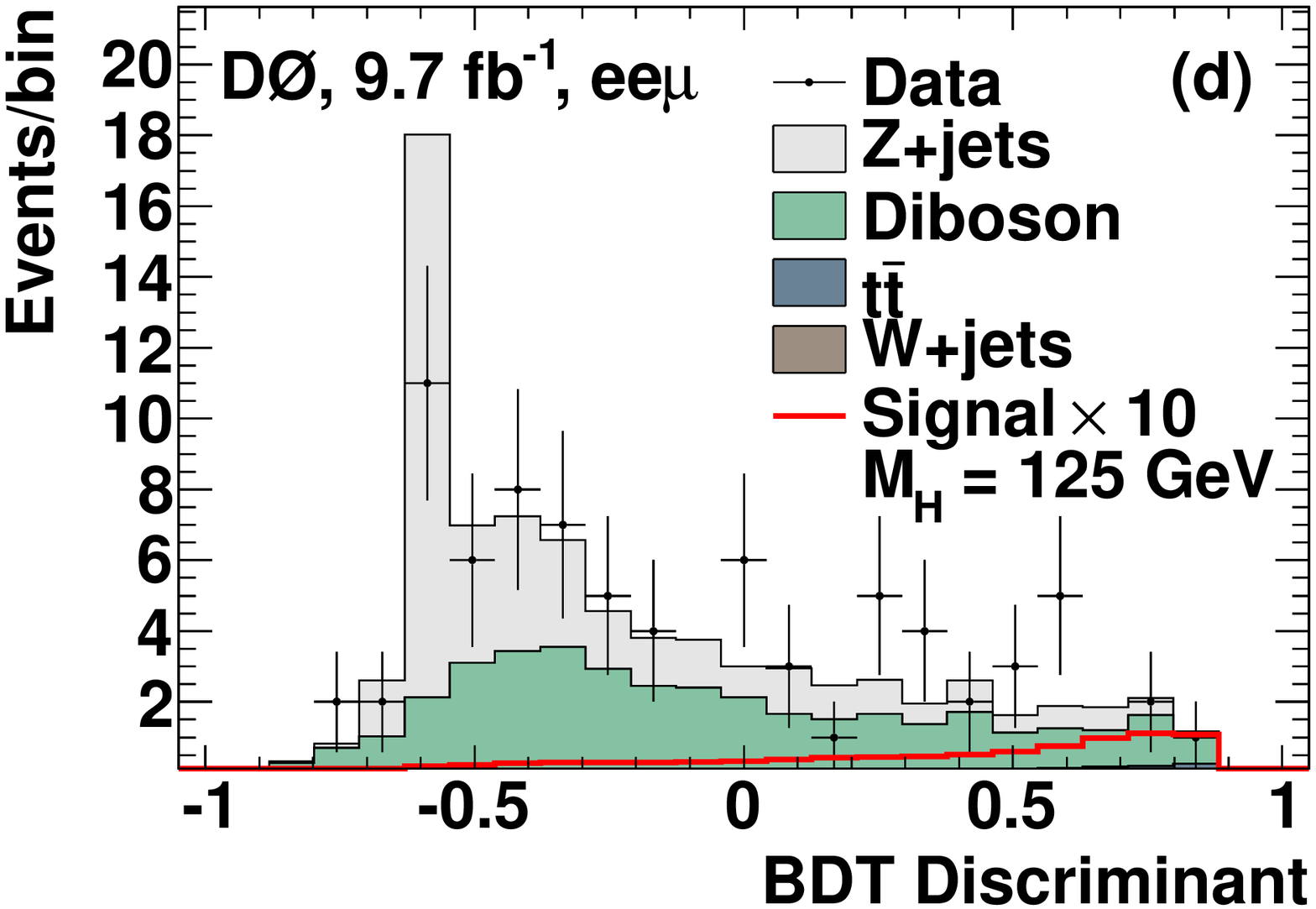}
\includegraphics[scale=0.295]{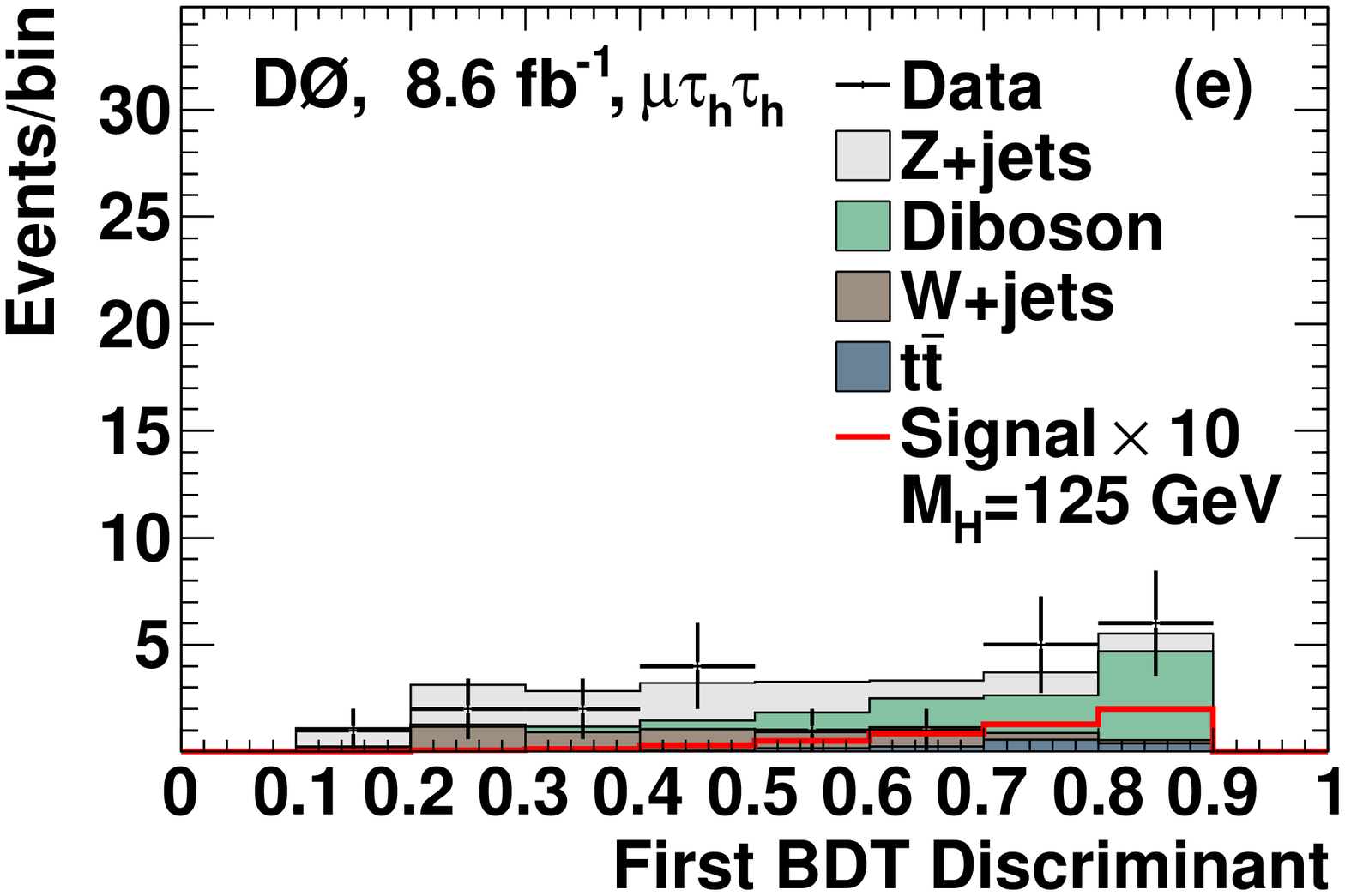}
\includegraphics[scale=0.295]{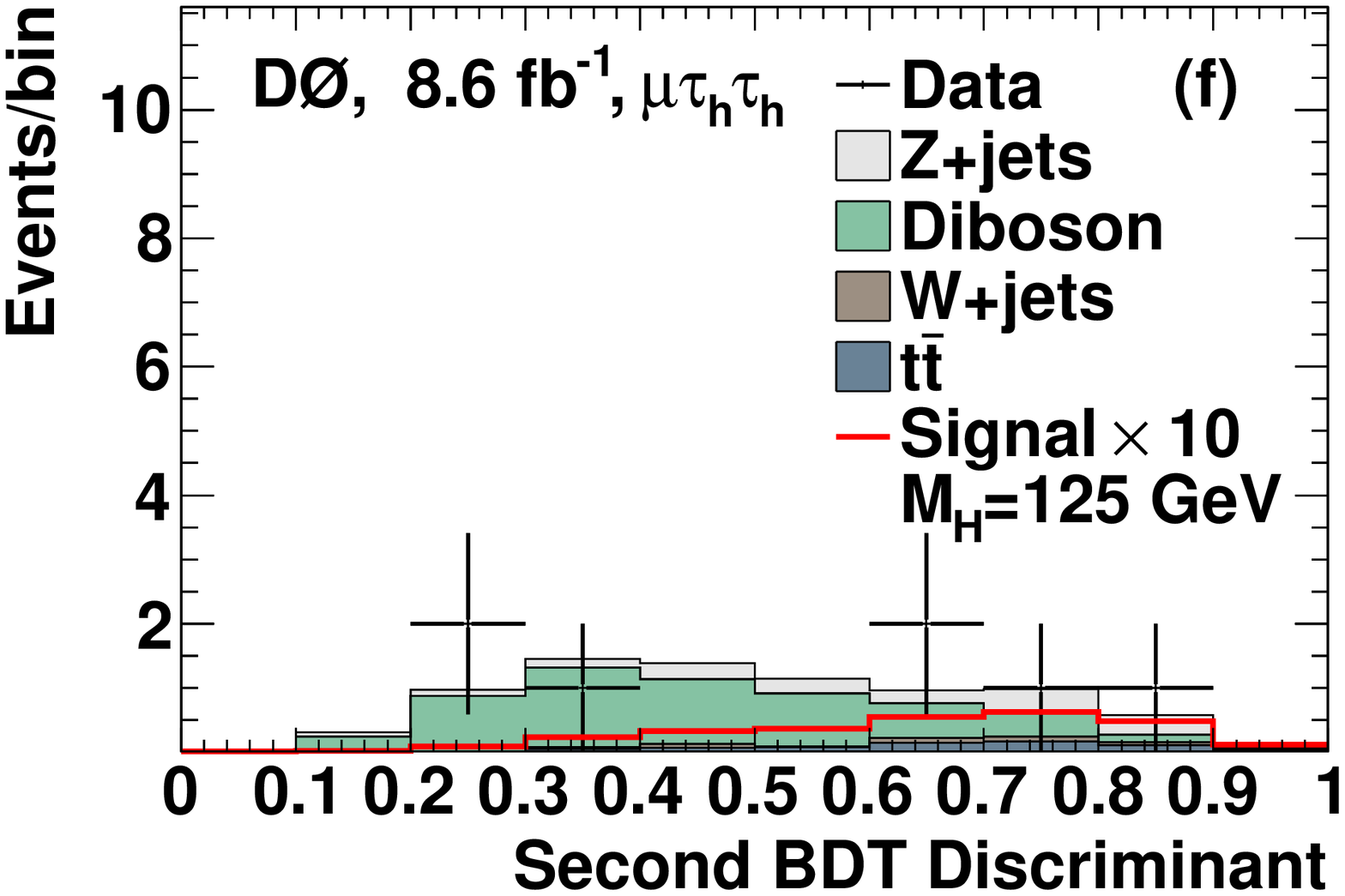}\\
\includegraphics[scale=0.295]{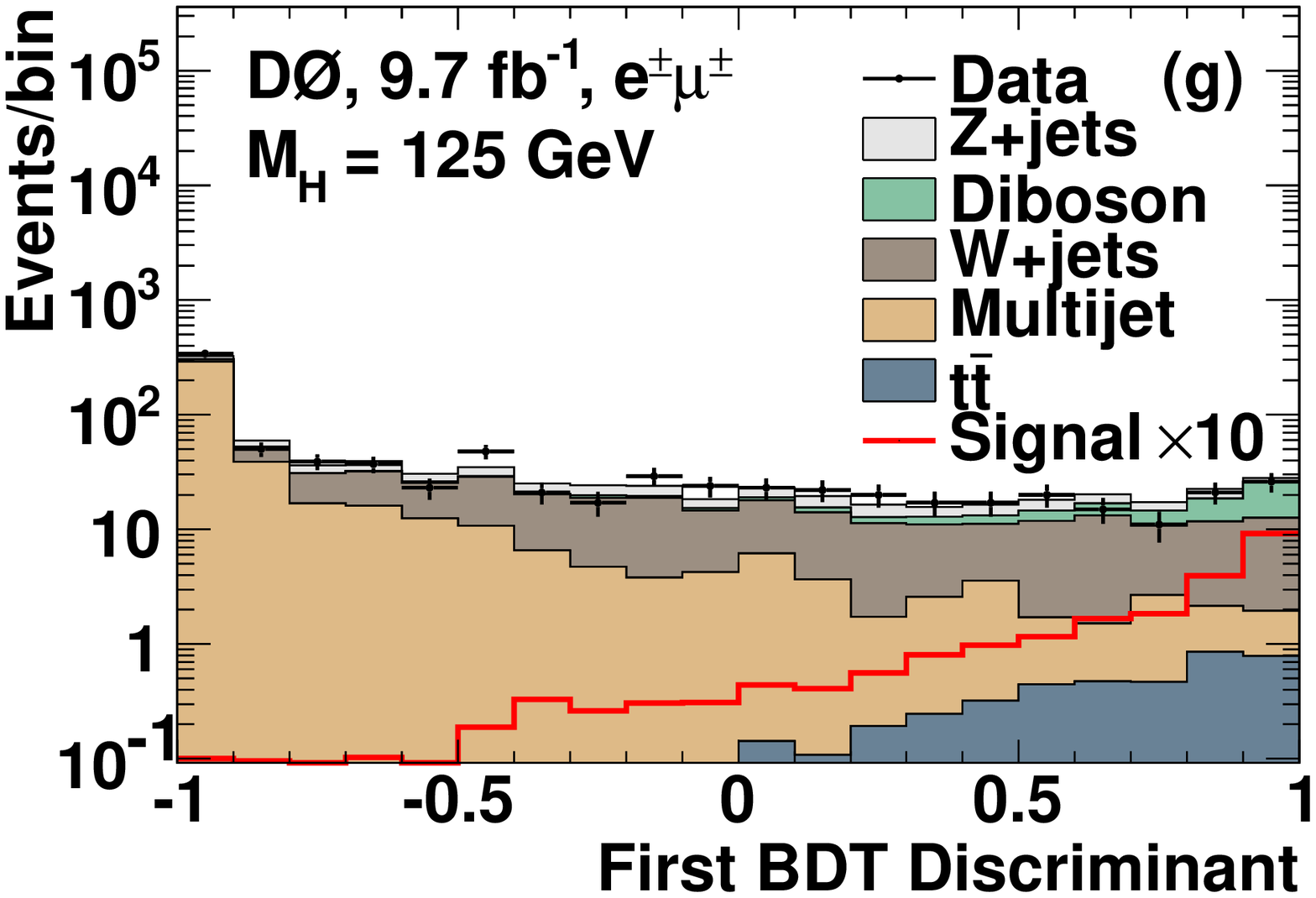} 
\includegraphics[scale=0.295]{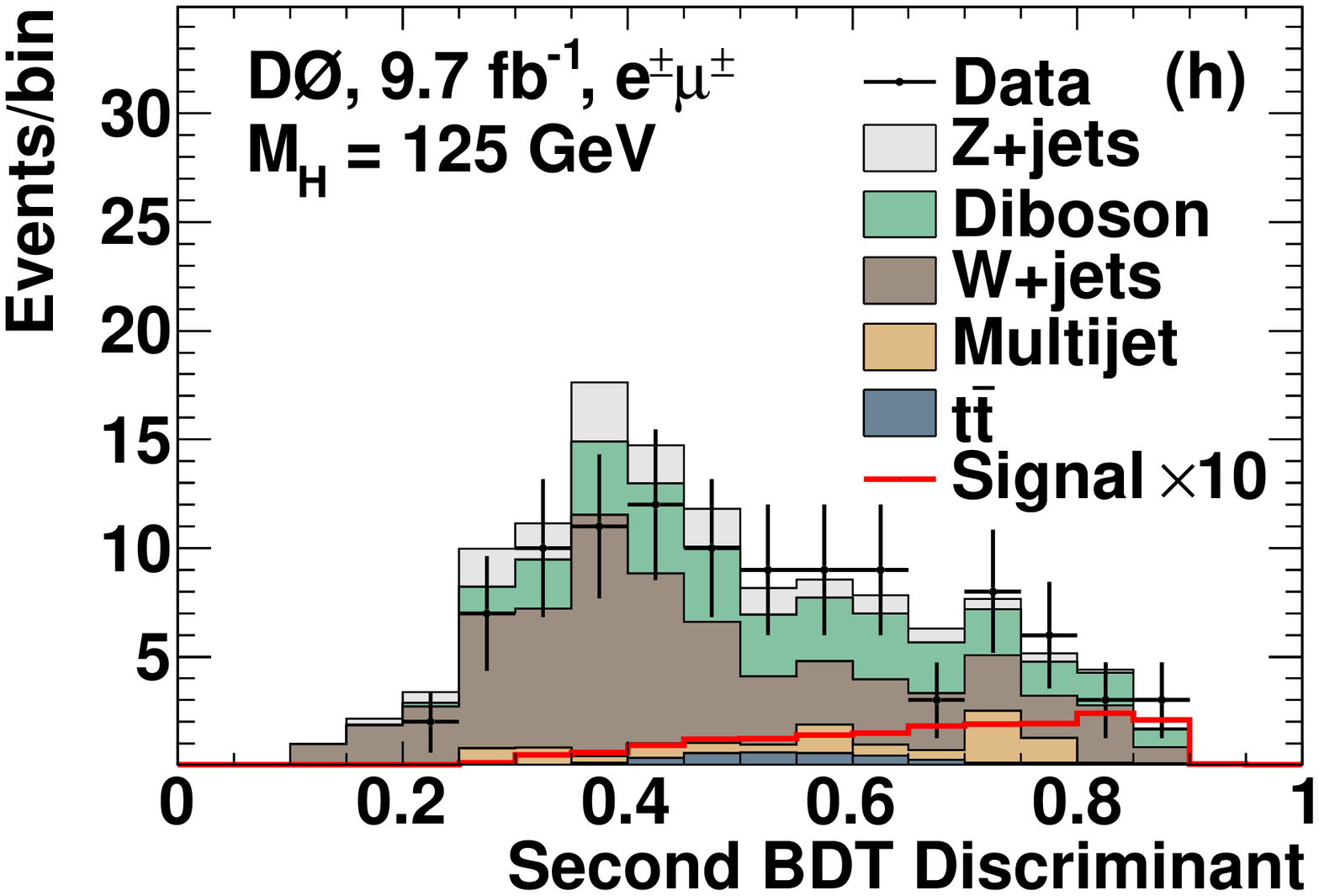}
\caption{\protect\label{fig-BDT} (color online).
Distribution of the BDT outputs for the (a) $e\mu\mu_{\rm A}$, (b) $e\mu\mu_{\rm B}$, 
(c) $e\mu\mu_{\rm C}$,  and (d) $ee\mu$ channels and
distributions of the outputs of the (e,g) first and (f,h) second BDT in the $\mu\tau_h\tau_h$
and $e^{\pm}\mu^{\pm}$ channels, respectively.                             
The data are compared to the sum of the expected background and
to simulations of a Higgs boson signal for $M_H=125$~GeV,
multiplied                                                                                                                                                                                                                         
by factors of 75 for the $e\mu\mu_{\rm B}$ and $e\mu\mu_{\rm C}$ channels and $10$ for the other channels.
}
\end{figure*}

A single BDT for each mass point is used to discriminate between signal and all
background processes for the $ee \mu$ sample and for
each of the three $e\mu\mu$ sub-samples. 
The output distributions of the BDTs, shown in 
Fig.~\ref{fig-BDT}(a)-(d) for data, signal with $M_H=125$~GeV, and 
for expected background, are used
to discriminate between signal and background.

Two BDTs are trained for the $\mu\tau_h\tau_h$ sample, where
the first BDT discriminates between signal and all background sources 
except diboson production and the second BDT between signal and the dominant
diboson background.
Events that pass a selection requirement on the first BDT discriminant 
of $>0.680$--$0.788$, determined separately for each $M_H$ value to optimize the discrimination
between signal and background,  are used as input to the second BDT. 
The output distributions of the first BDT with $M_H=125$~GeV
is shown in Fig.~\ref{fig-BDT}(e) and for the second BDT in Fig.~\ref{fig-BDT}(f), using
all events where the first BDT output is $>0.744$.
The output of the first BDT is used as the discriminant in the limit setting for all 
events that fail the requirement on the first BDT output and the output of the 
second BDT for all remaining events. 

The output distribution for the first BDT used for the $e^{\pm}\mu^{\pm}$ 
channel is shown in Fig.~\ref{fig-BDT}(g). It discriminates
mainly between signal and $W$+jets as well as multijet production.
After requiring the output of the first BDT to be $>0.3$ and $\min\{M_T(e),M_T(\mu)\}>7$~GeV
in a final selection step,
the number of expected background events at $M_H=125$~GeV
is reduced from $809\pm 93$ to $122\pm 7$, while
the expected number of signal events 
is only reduced by $23\%$ (see Table~\ref{tab:cutflow2}).
A second BDT is trained to discriminate between signal and the remaining
background sources, which are mainly from diboson, $W+$jets, and $Z+$jets production,
in the remaining events.
The output of the second BDT for this sample, shown in Fig.~\ref{fig-BDT}(h) for 
$M_H=125$~GeV, is
used as discriminant in the Higgs boson searches.
  
\begin{figure*}[htbp]
\includegraphics[scale=0.291]{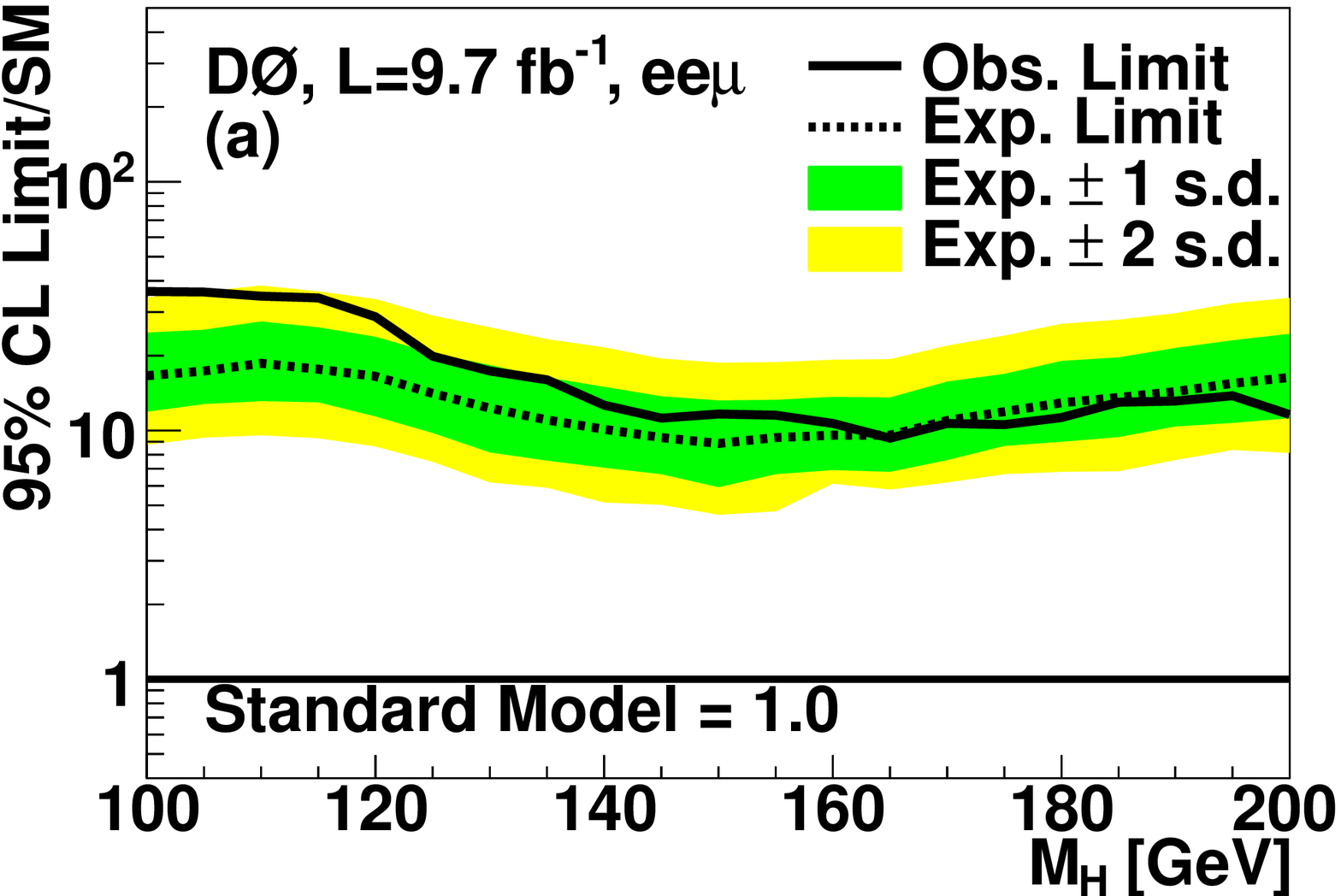}
\includegraphics[scale=0.295]{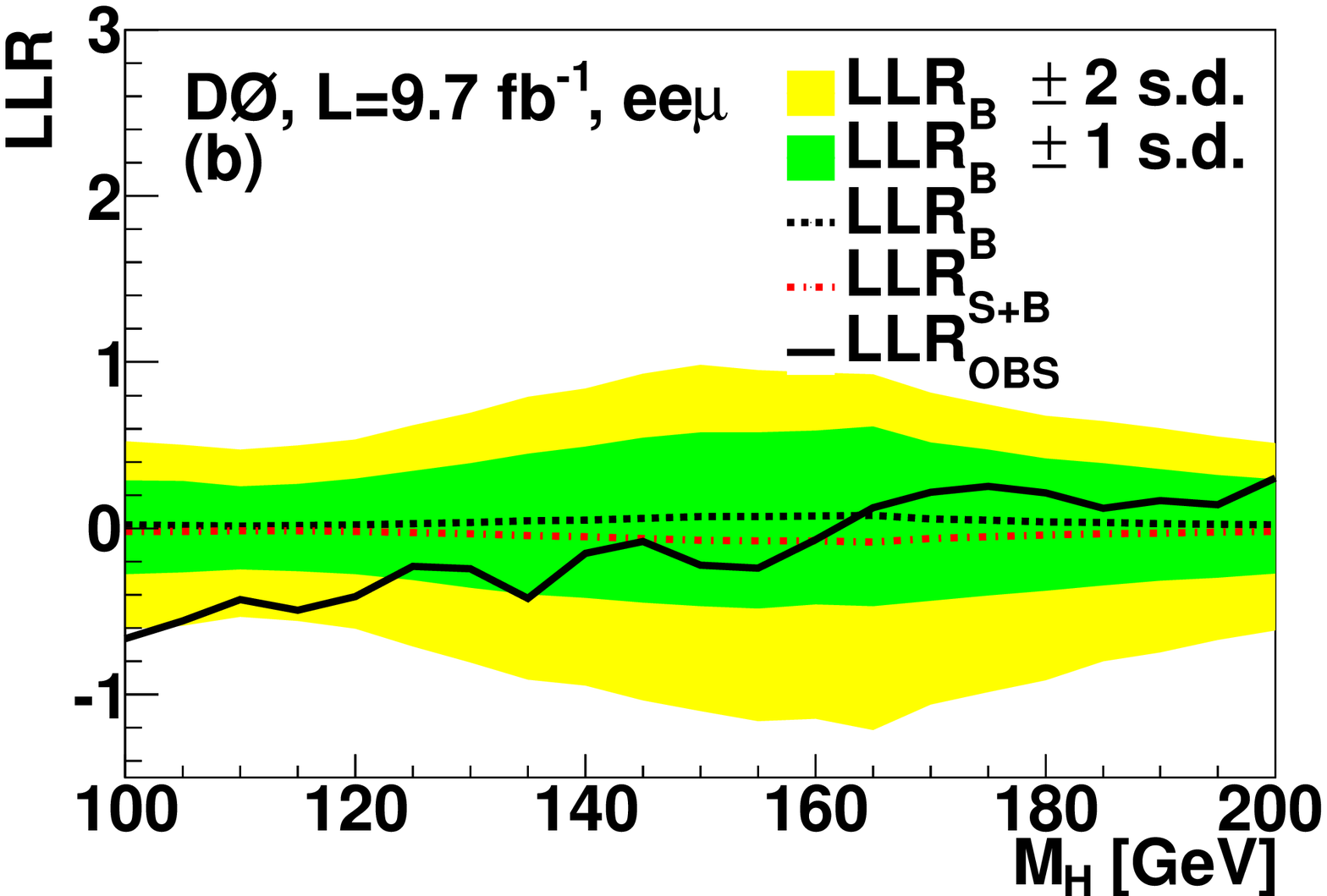}\\
\includegraphics[scale=0.291]{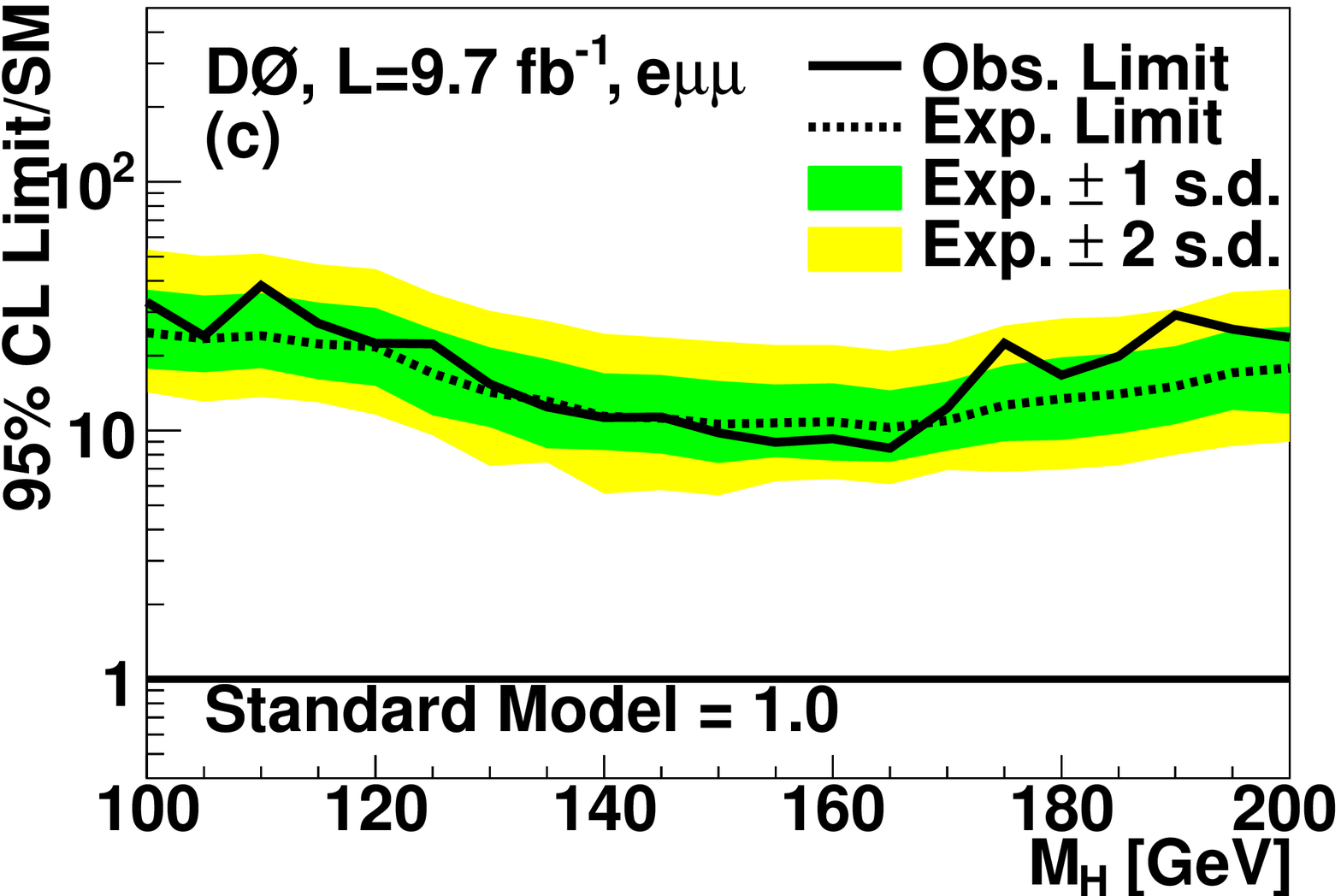}
\includegraphics[scale=0.295]{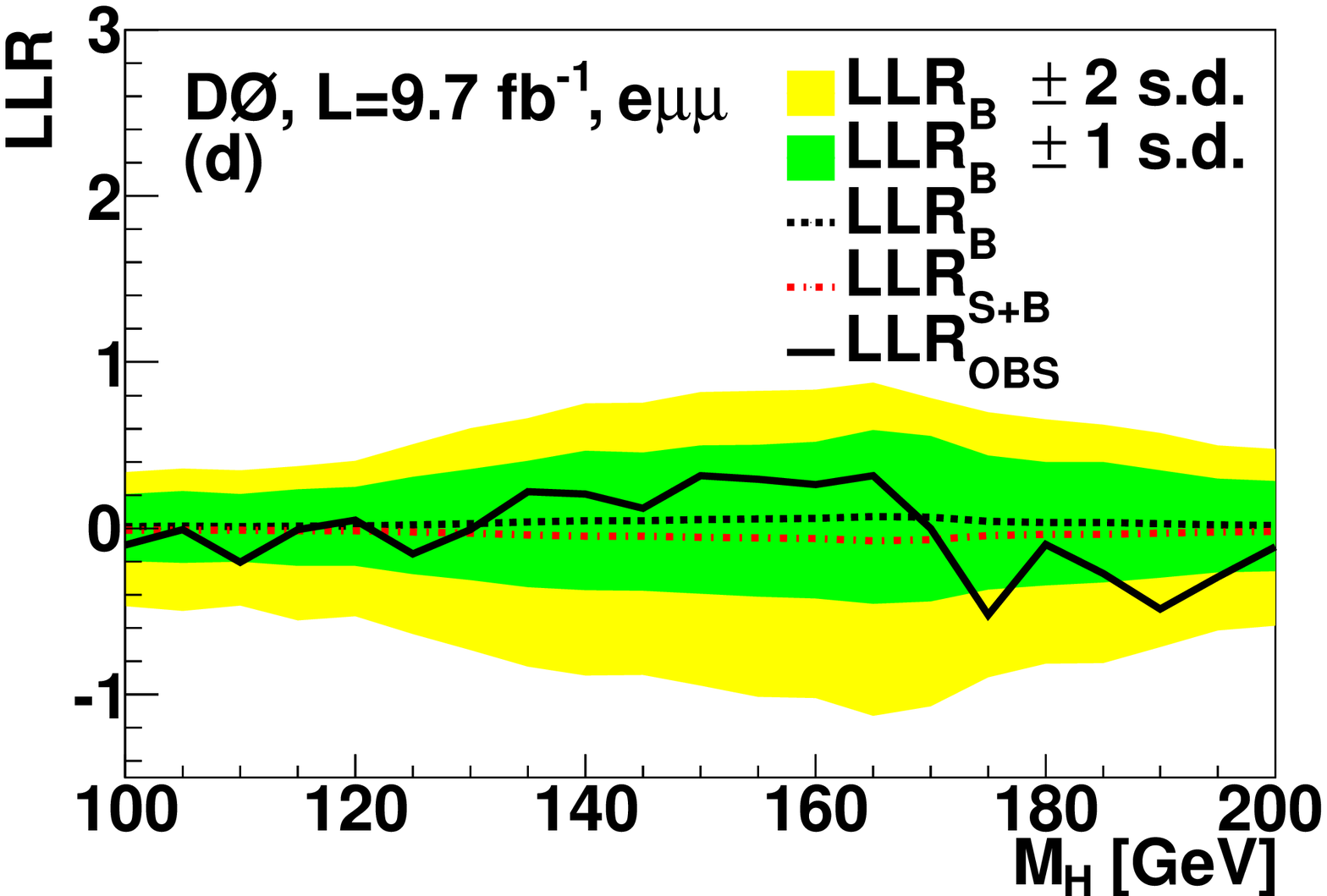}\\
\includegraphics[scale=0.291]{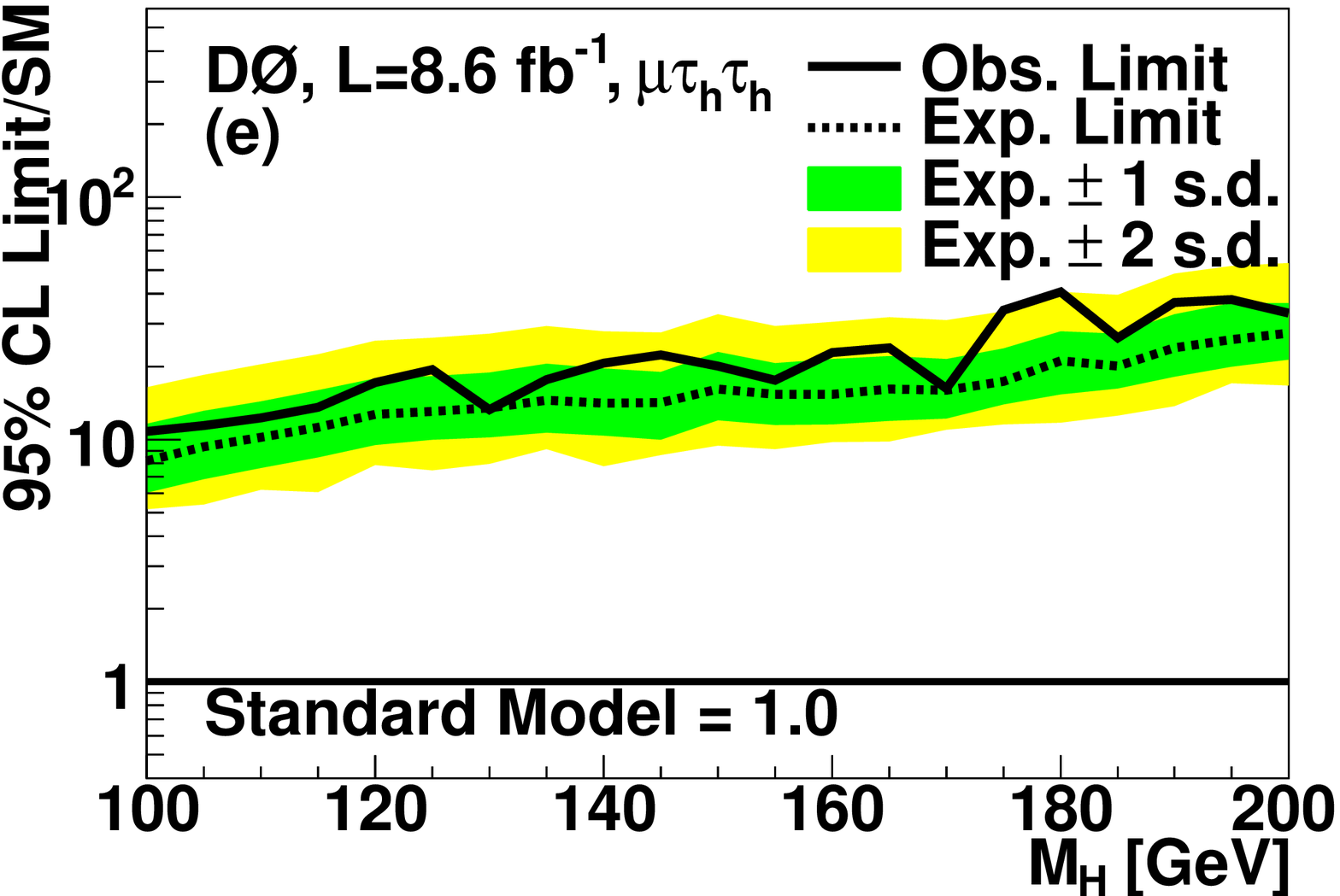}
\includegraphics[scale=0.295]{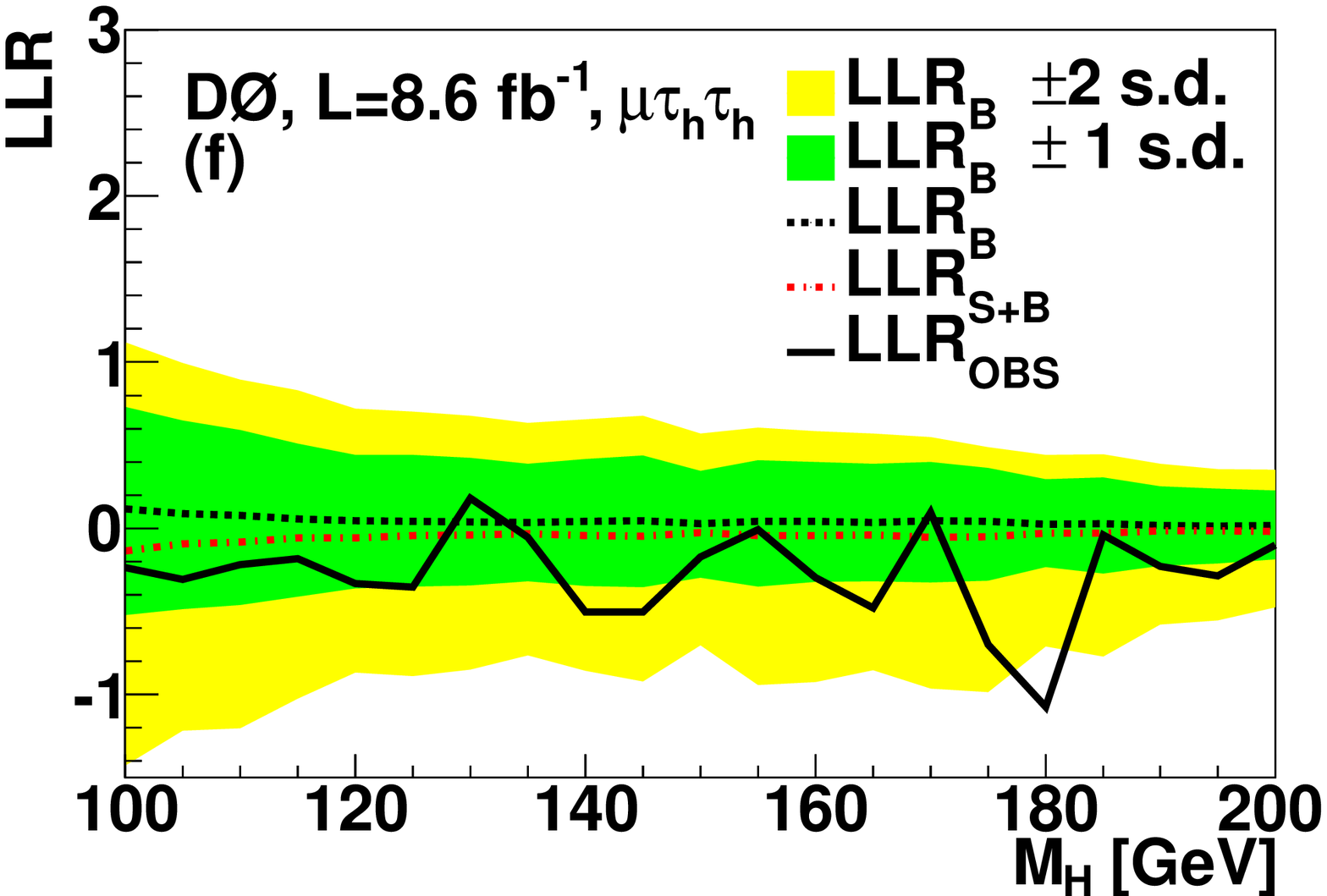}\\
\includegraphics[scale=0.291]{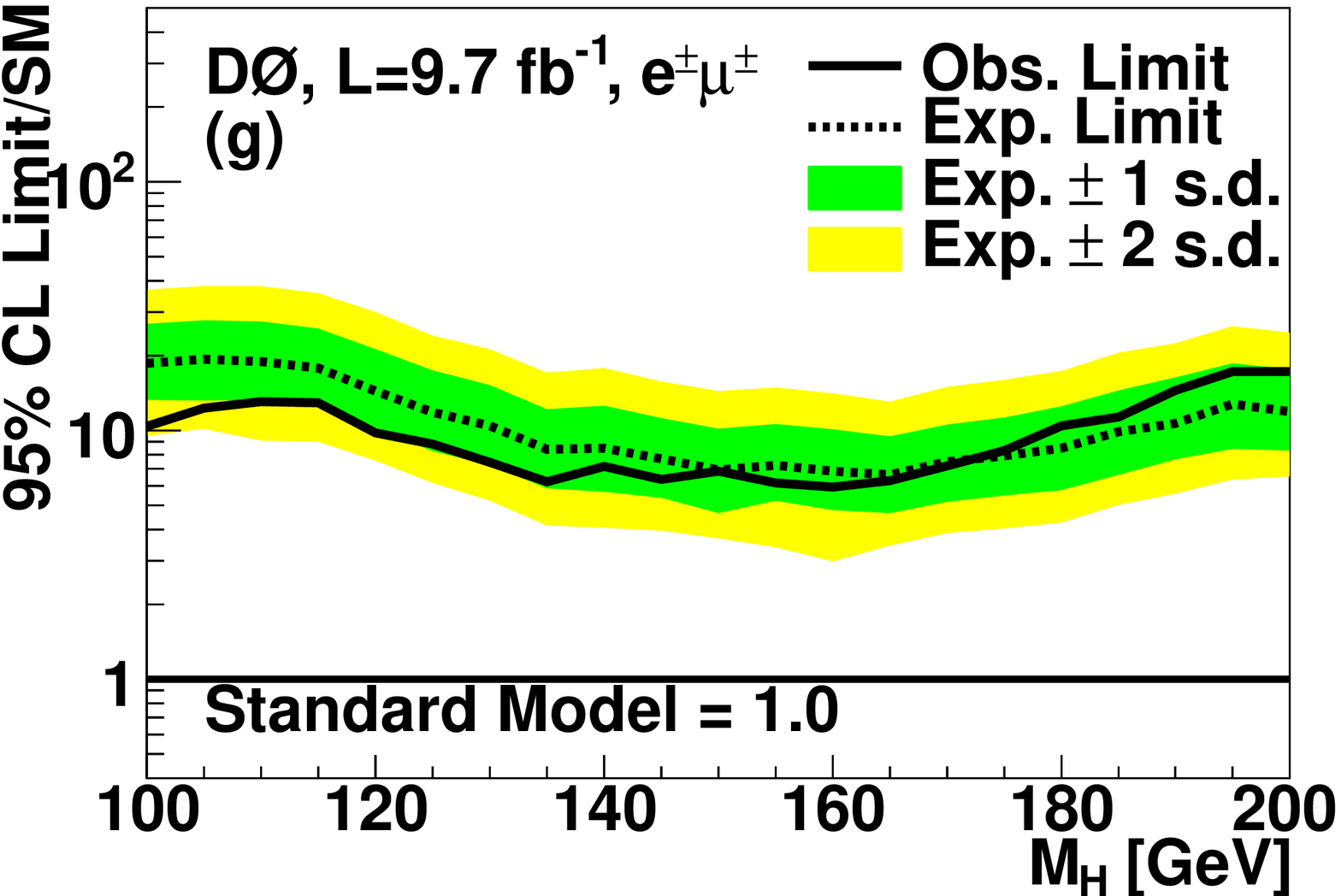}
\includegraphics[scale=0.295]{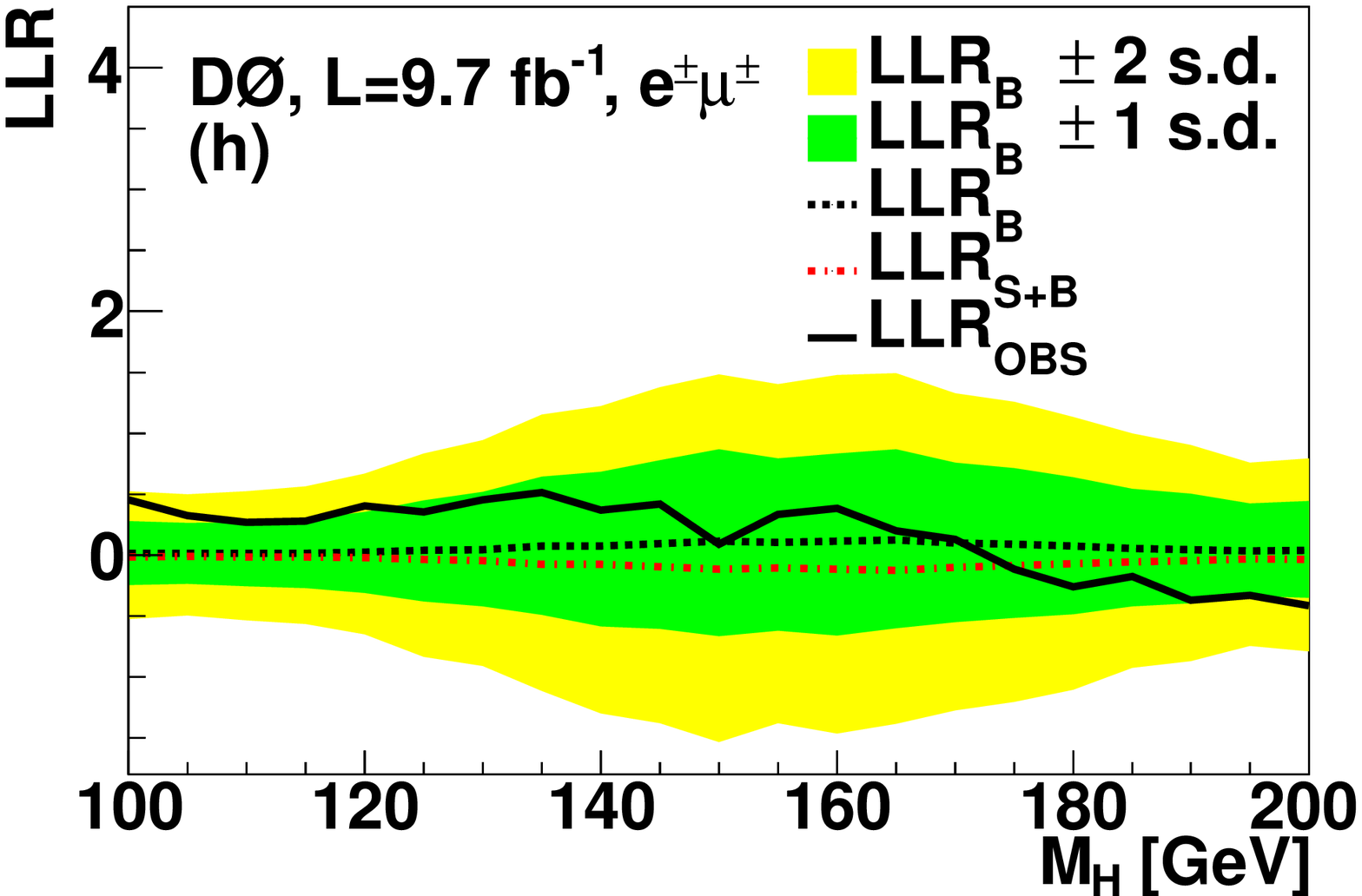}\\
\includegraphics[scale=0.291]{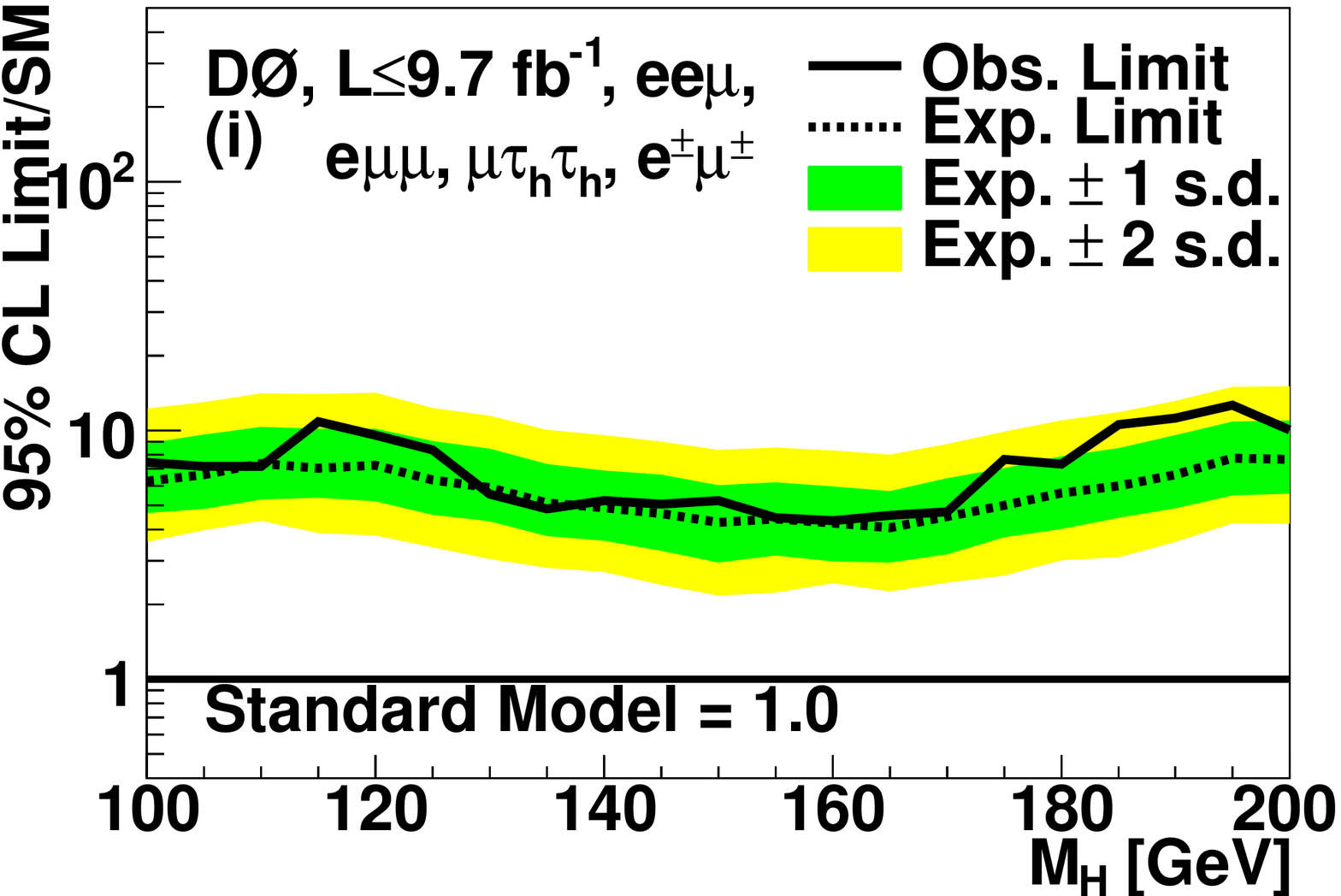}
\includegraphics[scale=0.295]{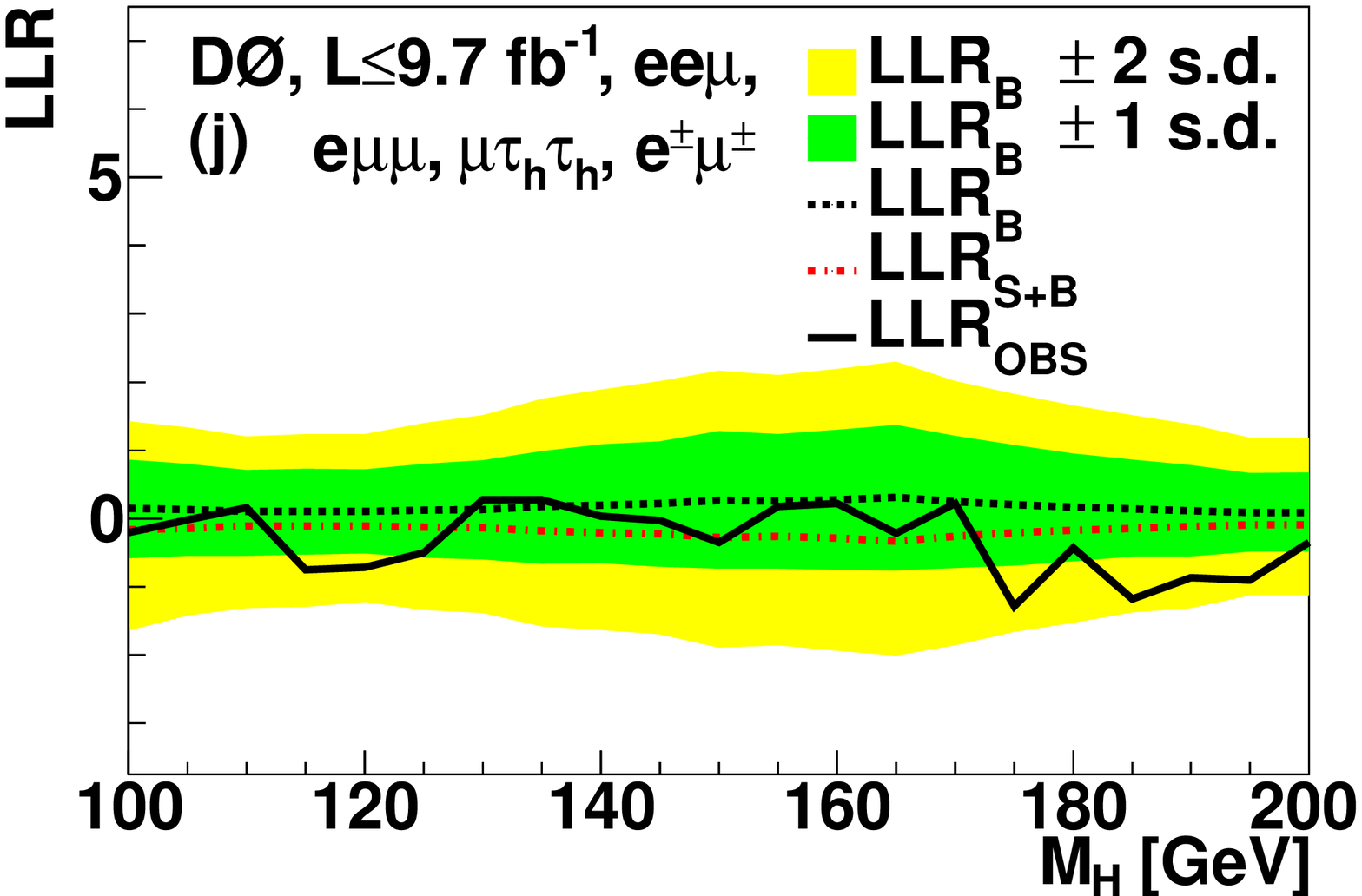}
\caption{\protect\label{fig:smlimits} (color online).
Upper limit on the SM Higgs boson production cross section expressed
as a ratio to the SM prediction (left column) and observed
LLR (right column) as a function of $M_H$ for the (a,b) $ee \mu$, (c,d) $e\mu\mu$, (e,f) $\mu\tau_h\tau_h$, (g,h) $e^{\pm} \mu^{\pm}$ channels, and (i,j) for all channels combined. 
The LLRs are shown for the background-only and the signal-plus-background hypotheses.
The bands correspond to regions of $\pm 1$ and $\pm 2$ standard deviations (s.d.) around 
the median expected limit and around the 
expected median LLR for the background-only hypothesis, respectively.
}
\end{figure*}
\begin{figure}[htb]
\includegraphics[scale=0.291]{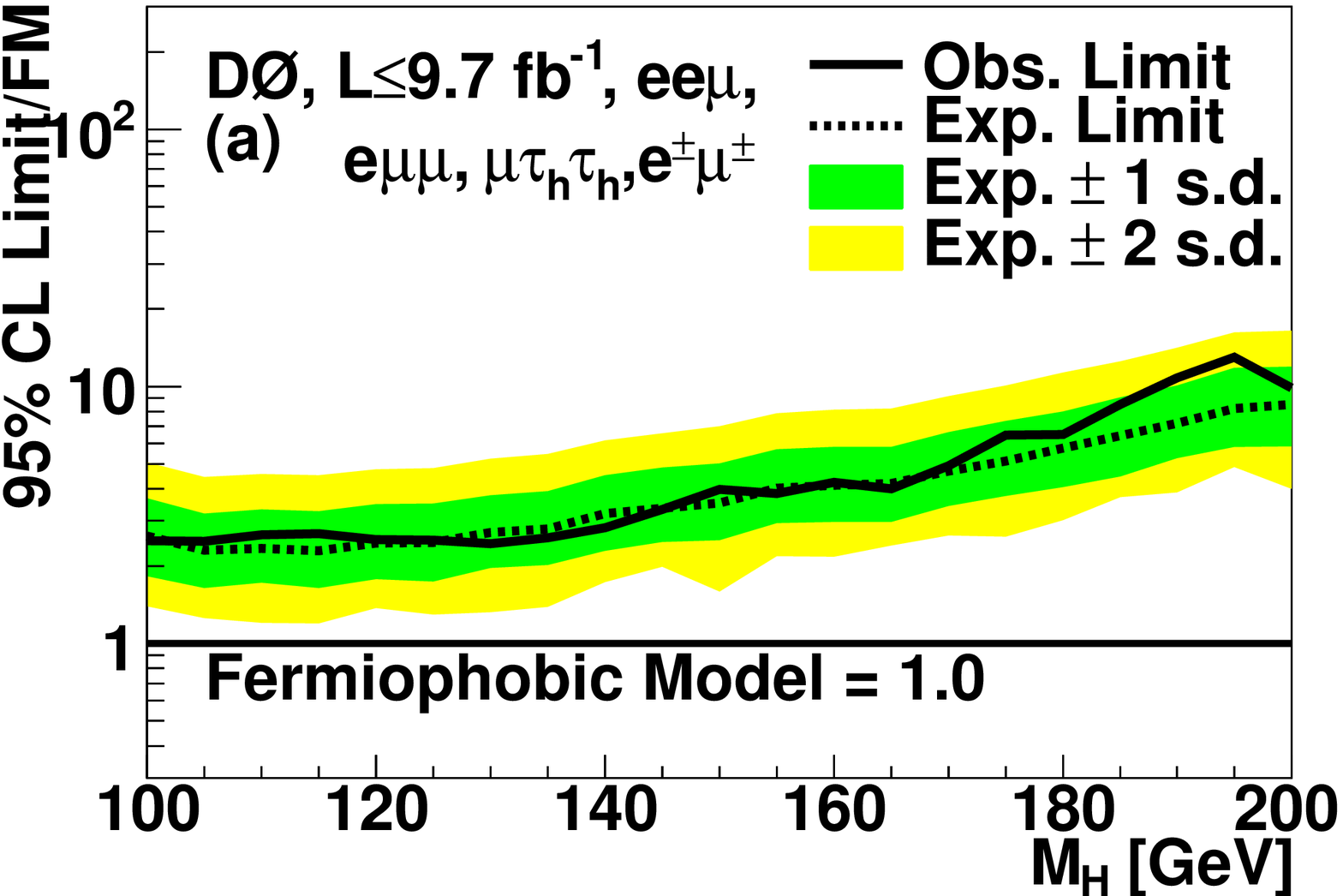}
\includegraphics[scale=0.295]{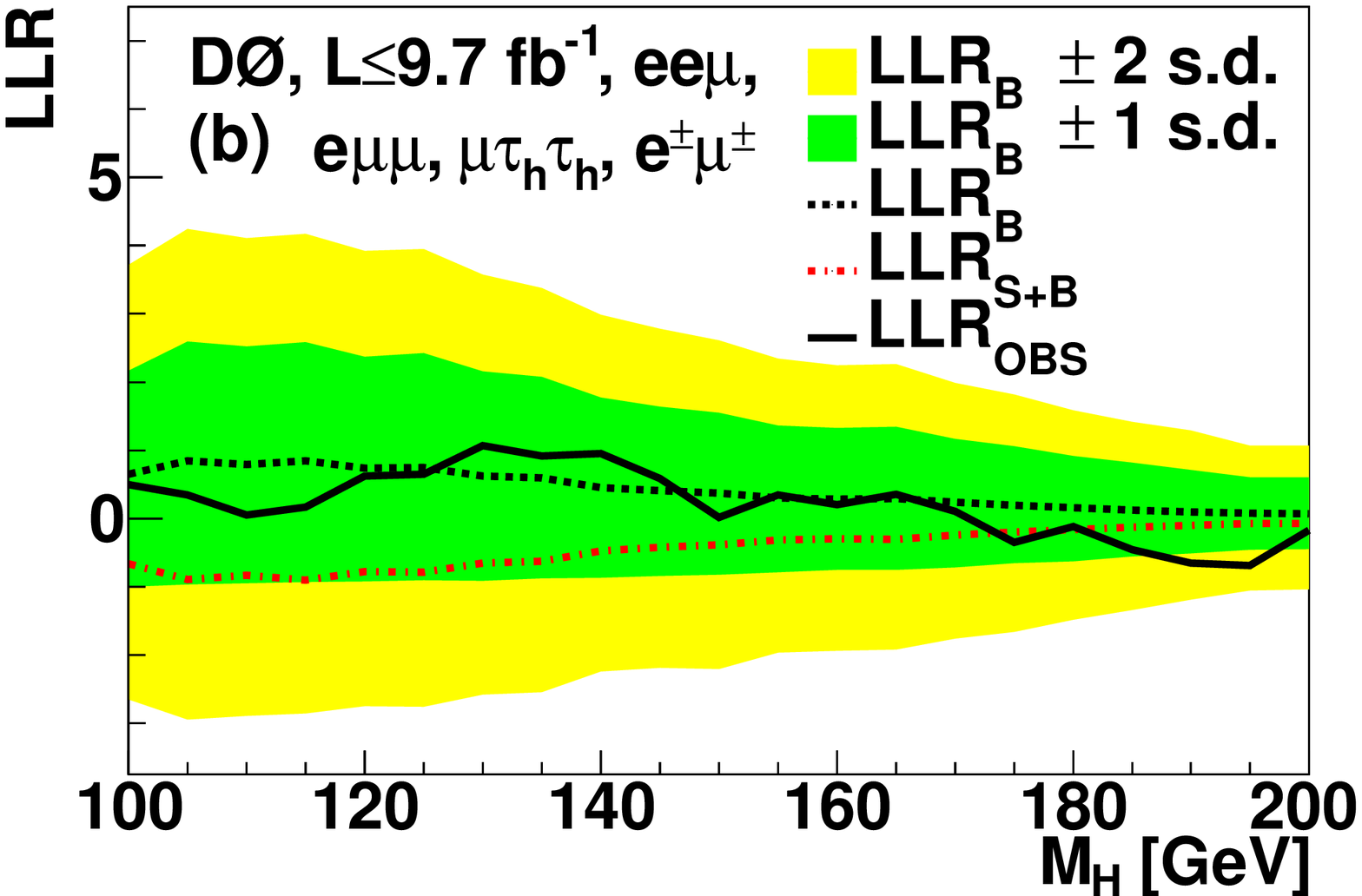}
\caption{\protect\label{fig:fllr} (color online).
(a) Upper limit on the fermiophobic Higgs boson production cross section expressed
 as a ratio to the prediction and (b) 
 observed LLR as a function of $M_H$ for the combined channels.
 Also shown are the expected LLR distributions for the background-only hypothesis and
for the signal+background hypothesis, with the bands indicating $\pm 1$ and $\pm2$ s.d.~fluctuations 
around the expected median LLR for the background-only hypothesis.
}
\end{figure}

\section{Systematic Uncertainties}

Systematic uncertainties on both background and signal, 
including their correlations, are taken into account as uncertainties
on the normalizations and on the shape of differential distributions.
The theoretical uncertainty on the cross sections for 
$Z/\gamma^{*}\to\ell^+\ell^-$, 
$W$+jets, $t\bar{t}$, and diboson production vary between $6\%$ and $7\%$.
The theoretical uncertainty on the $W$+jets cross section is 
applied for the $ee\mu$ and $e\mu\mu$ channels, where this background is 
normalized using the theoretical prediction. 
The uncertainty from the normalization of the $W$+jets background in the
in the $\mu\tau_h\tau_h$ and $e^{\pm}\mu^{\pm}$ channels is $5\%-6\%$.
Additional shape dependent uncertainties on the $W$+jets distributions
are applied in the $e^{\pm}\mu^{\pm}$ channel to account for uncertainties
in the description of the $p_T$ spectrum of $W$ bosons as well as initial and final state
radiation.
The systematic uncertainty on the normalization of the multijet background
in the $e^{\pm}\mu^{\pm}$ channel is estimated to be $20\%$ by studying its
dependence on jet multiplicity and lepton $p_T$.

The theoretical uncertainty on signal production cross sections is 5\% for 
gluon-gluon fusion and $6.2\%$ for associated {\sl VH} production.
The uncertainty on the measured integrated luminosity is $6.1\%$~\cite{bib-lumi}.
The systematic uncertainty on the lepton identification efficiency 
is $1.8\% - 4\%$ per muon and $2.5\%$ per electron.  
The total uncertainty on the identification efficiency for 
both $\tau_h$ candidates, including the uncertainty on the neural network discriminant
used to distinguish $\tau_h$ candidates from jets, is $7\%$ per event. 
An uncertainty on the normalization of the signal and
the simulated background is derived by comparing distributions of
data obtained using the inclusive trigger method with
samples obtained using only single-lepton triggers. 
The resulting uncertainty is $1.5\% - 5\%$.

\begin{table}[htbp]
\caption{\label{tab:smlimits}
Expected and observed $95\%$~C.L.~upper 
limits on the SM Higgs boson production cross section
relative to the SM expected value, for the $ee\mu$, $e\mu\mu$, $\mu\tau_h\tau_h$, and 
$e^{\pm}\mu^{\pm}$  channels 
separately and combined.
}\begin{center}
\begin{tabular}{c|cc|cc|cc|cc|cc} 
\hline
\hline
$M_H$ & \multicolumn{2}{c|}{$ee\mu$} & \multicolumn{2}{c|}{$e\mu\mu$} & 
\multicolumn{2}{c|}{$\mu\tau_h\tau_h$} & \multicolumn{2}{c|}{$e^{\pm}\mu^{\pm}$} &
\multicolumn{2}{c}{Combined} \\ \cline{2-11}
(GeV) & Exp & Obs &  Exp & Obs &  Exp & Obs & Exp & Obs & Exp & Obs  \\ 
\hline
100  &    16.6 &    36.1 &    24.8 &    32.9 &    8.2  &    10.8 &    18.6 &    10.4 &    6.3
  &    7.5     \\
105  &    17.4 &    36.1 &    23.5 &    24.0 &    9.3  &    11.4 &    19.3 &    12.3 &    6.7
  &    7.2     \\
110  &    18.6 &    34.8 &    24.0 &    38.2 &    10.2 &    12.3 &    18.9 &    13.0 &    7.4
  &    7.2     \\
115  &    17.7 &    34.1 &    22.3 &    27.1 &    11.3 &    13.6 &    17.8 &    12.9 &    7.1
  &    10.9    \\
120  &    16.5 &    28.6 &    21.7 &    22.5 &    12.7 &    17.2 &    14.4 &    9.8  &    7.3
  &    9.6     \\
125  &    14.1 &    19.9 &    17.0 &    22.3 &    13.0 &    19.4 &    11.8 &    8.8  &    6.3
  &    8.4     \\
130  &    12.3 &    17.4 &    14.3 &    15.4 &    13.5 &    13.3 &    10.4 &    7.4  &    5.9
  &    5.5     \\
135  &    11.0 &    16.0 &    13.1 &    12.4 &    14.6 &    17.6 &    8.4  &    6.2  &    5.1
  &    4.9     \\
140  &    10.1 &    12.6 &    11.4 &    11.3 &    14.1 &    20.6 &    8.5  &    7.2  &    4.9
  &    5.2     \\
145  &    9.4  &    11.2 &    11.2 &    11.4 &    14.2 &    22.3 &    7.7  &    6.4  &    4.6
  &    5.1     \\
150  &    8.9  &    11.7 &    10.6 &    9.8  &    16.2 &    20.1 &    7.0  &    6.9  &    4.3
  &    5.2     \\
155  &    9.3  &    11.5 &    10.8 &    9.0  &    15.4 &    17.6 &    7.3  &    6.2  &    4.4
  &    4.5     \\
160  &    9.6  &    10.7 &    10.9 &    9.3  &    15.4 &    22.8 &    6.9  &    5.9  &    4.2
  &    4.4     \\
165  &    9.6  &    9.3  &    10.3 &    8.5  &    16.1 &    23.9 &    6.6  &    6.3  &    4.1
  &    4.6     \\
170  &    11.0 &    10.7 &    11.0 &    12.3 &    16.0 &    16.2 &    7.5  &    7.2  &    4.5
  &    4.7     \\
175  &    11.9 &    10.6 &    12.7 &    22.4 &    17.4 &    34.3 &    7.9  &    8.3  &    5.0
  &    7.7     \\
180  &    12.9 &    11.3 &    13.5 &    16.7 &    21.1 &    40.7 &    8.5  &    10.4 &    5.6
  &    7.3     \\
185  &    13.6 &    13.0 &    14.1 &    19.8 &    20.1 &    26.2 &    9.9  &    11.3 &    6.0
  &    10.6    \\
190  &    14.4 &    13.1 &    15.1 &    29.1 &    24.0 &    36.7 &    10.7 &    14.5 &    6.7
  &    11.2    \\
195  &    15.5 &    13.8 &    17.1 &    25.6 &    25.9 &    37.8 &    12.8 &    17.3 &    7.7
  &    12.7    \\
200  &    16.4 &    11.6 &    17.8 &    23.7 &    27.5 &    33.3 &    11.9 &    17.3 &    7.7
  &    10.1    \\
\hline \hline
\end{tabular}
\end{center}
\end{table}

The uncertainty
on the signal acceptance from the uncertainty on the parton distribution functions is~$2.5\%$.
The uncertainty on the probability that leptons originating from jets are 
selected in the $ee\mu$ and $e\mu\mu$ 
channels is $30\%$ for the $Z/W$+jets and the $WW$ samples.  
The uncertainty on charge misidentification is $20\%$ for the $e\mu\mu$ channel and
$16\%$ for the $Z$+jets background and $50\%$ for the $\ttbar$, and {\sl WW} 
background in the $e^{\pm} \mu^{\pm}$ channel. 
 A $7.3\%$ systematic uncertainty is assessed on the $Z\gamma$ background 
in the $e\mu\mu$ final state.  
The uncertainties on the description of the $p_T$ distributions
of $W$ and $Z$ boson, the uncertainties on the $p_T$ resolutions for electrons and muons,
and the uncertainties from jet energy resolution and efficiencies 
are found to have a negligible effect on the results.

\begin{table}[htbp]
\begin{center}
\begin{tabular}{c|cc|cc|cc|cc|cc} 
\hline
\hline
$M_H$ & \multicolumn{2}{c|}{$ee\mu$} & \multicolumn{2}{c|}{$e\mu\mu$} & \multicolumn{2}{c|}{$\mu\tau_h\tau_h$} & \multicolumn{2}{c|}{$e^{\pm}\mu^{\pm}$} & \multicolumn{2}{c}{Combined} \\ \cline{2-11}
(GeV) & Exp & Obs &  Exp & Obs &  Exp & Obs & Exp & Obs &  Exp & Obs \\ \hline
100  &    5.1  &    10.4 &    5.7  &    7.6  &    26 &    27 &    3.7  &    2.1  &    2.6
  &    2.5     \\
105  &    4.4  &    9.5  &    5.1  &    5.0  &    17 &    16 &    3.7  &    2.4  &    2.3
  &    2.5     \\
110  &    5.3  &    9.2  &    5.4  &    9.1  &    16 &    13 &    3.4  &    2.3  &    2.3
  &    2.7     \\
115  &    4.6  &    8.6  &    4.9  &    6.2  &    15 &    15 &    3.8  &    2.7  &    2.3
  &    2.7     \\
120  &    5.1  &    8.5  &    5.8  &    6.5  &    17 &    25 &    3.7  &    2.5  &    2.5
  &    2.5     \\
125  &    5.1  &    7.0  &    5.8  &    7.7  &    17 &    18 &     3.7  &    2.8  &    2.5
  &    2.5     \\
130  &    5.5  &    7.5  &    6.1  &    6.6  &    18 &    17 &    4.1  &    3.0  &    2.7
  &    2.4     \\
135  &    6.0  &    8.3  &    6.9  &    6.1  &    17 &    20 &    4.1  &    3.1  &    2.8
  &    2.6     \\
140  &    6.7  &    8.3  &    7.2  &    6.8  &    17 &    18 &    5.1  &    4.1  &    3.2
  &    2.8     \\
145  &    7.2  &    8.3  &    8.1  &    8.3  &    16 &    20 &    5.3  &    4.5  &    3.4
  &    3.3     \\
150  &    7.7  &    9.8  &    8.9  &    7.6  &    20 &    28 &    5.6  &    5.4  &    3.5
  &    4.0     \\
155  &    8.4  &    10.7 &    9.7  &    7.9  &    21 &    23 &    6.5  &    5.4  &    4.0
  &    3.8     \\
160  &    9.4  &    10.5 &    10.4 &    9.0  &    19 &    30 &    6.6  &    5.7  &    4.1
  &    4.2     \\
165  &    9.4  &    9.3  &    10.0 &    8.3  &    22 &    32 &    6.6  &    6.1  &    4.2
  &    4.0     \\
170  &    11.2 &    10.8 &    11.2 &    12.1 &    23 &    30 &    7.3  &    7.0  &    4.7
  &    4.9     \\
175  &    12.1 &    10.6 &    12.7 &    21.9 &    25 &    29 &    7.9  &    8.2  &    5.1
  &    6.5     \\
180  &    13.5 &    11.4 &    14.4 &    16.4 &    29 &    33 &    8.5  &    10.1 &    5.8
  &    6.5     \\
185  &    14.8 &    13.7 &    15.3 &    20.6 &    31 &    45 &    10.0 &    11.4 &    6.4
  &    8.5     \\
190  &    16.3 &    13.5 &    17.0 &    30.5 &    35 &    50 &    11.0 &    14.4 &    7.2
  &    10.8    \\
195  &    17.8 &    15.5 &    19.7 &    29.1 &    40 &    62 &    12.9 &    17.3 &    8.2
  &    13.0    \\
200  &    19.0 &    12.0 &    20.6 &    23.9 &    43 &    47 &    12.1 &    17.3 &    8.5
  &    9.9     \\
\hline \hline
\end{tabular}
\end{center}
\caption{\label{tab:flimits}
Expected and observed $95\%$~C.L.~upper 
limits on the fermiophobic Higgs boson production cross section
relative to the expected cross section in the fermiophobic model,
for the $ee\mu$, $e\mu\mu$, $\mu\tau_h\tau_h$, and 
$e^{\pm}\mu^{\pm}$  channels 
separately and combined.
}
\end{table}

\section{Results on the SM Higgs boson}

We determine limits on the SM Higgs boson production cross section 
using a modified frequentist approach~\cite{cls} using the distributions
of the final discriminants shown in Fig.~\ref{fig-BDT}.
A log-likelihood ratio (LLR) test statistic 
is formed using the Poisson probabilities for estimated
background yields, the signal acceptance, and the observed
number of events for different Higgs boson mass hypotheses.
The confidence levels are derived by integrating the
LLR distribution in pseudo-experiments using both the signal-plus-background 
({\it CL}$_{{\rm sb}}$) and the background-only 
hypotheses ({\it CL}$_{\rm b}$). The excluded production cross section is
taken to be the cross section for which the confidence
level for signal, {\it CL}$_{\rm s}=${\it CL}$_{{\rm sb}}/${\it CL}$_{\rm b}$, equals $0.05$. 
The limits on the cross section for the different final states are given in Table~\ref{tab:smlimits}.  
The individual channels have similar sensitivity,
and the combined upper limits only vary within about a factor of two over the entire mass
range of $100 \le M_H< 200$~GeV. 
At $M_H=125$~GeV the expected and observed upper limits on the cross section, expressed
as a ratio relative
to the predicted SM cross section, are
6.3 and 8.4, respectively.
Figure~\ref{fig:smlimits} shows the limits on the cross section and the LLR distributions 
for each channel and for the combined result.

\section{Results on fermiophobic Higgs bosons}

In addition, we set limits in
a fermiophobic Higgs boson model, where the Higgs boson is assumed to couple to
$W$ and $Z$ bosons with SM strengths and 
the Higgs boson couplings to fermions are zero.
The gluon-gluon fusion Higgs production cross section is therefore small and is neglected.
The outputs of the BDTs trained using a SM Higgs boson signal are used to set the limits
with signal distributions where the gluon-gluon fusion processes and the $H\to\tau^+\tau^-$ decays have 
been removed and only the {\sl VH} 
production mechanism is considered. The limits on the cross section, given
as a ratio relative to the cross section in the fermiophobic
model, are listed in Table~\ref{tab:flimits}.
The $ee \mu$, $e\mu\mu$, and $e^{\pm}\mu^{\pm}$ channels have
similar sensitivity to a fermiophobic Higgs boson, whereas the
$\mu\tau_h\tau_h$ channel is less sensitive because $H\to\tau^+\tau^-$ 
decays do not contribute. The combined expected upper limits vary between $2.3$ and
$8.5$ and the observed limits between $2.4$ and $13.0$, expressed
as a ratio relative to the cross section in the fermiophobic model.
Figure~\ref{fig:fllr} shows the limits
on the cross section for the production of a fermiophobic Higgs boson and 
the LLR distribution.

\section{Conclusion}

We have presented the first search for the SM Higgs boson in 
multilepton $ee\mu$, $e\mu\mu$, and
$\mu\tau_h\tau_h$ final states and in like-charge $e^{\pm}\mu^{\pm}$ final states using
the D0 detector. The search is mainly sensitive to associated production of a $W$ or $Z$
boson with a Higgs boson, where the Higgs boson decays 
into $W^+W^-$ and $ZZ$ pairs, thereby 
probing the {\it HVV} coupling directly in production and decay.
We set limits on the cross section for a SM Higgs boson in the range $100 \le M_H \le 200$~GeV
with expected and observed upper limits of 
6.3 and 8.4 at $M_H=125$~GeV, expressed as the ratio relative to the predicted SM cross section.
We also interpret the data in a fermiophobic Higgs boson model.

\input acknowledgement.tex
\end{document}

%% file: author_list.tex
\affiliation{LAFEX, Centro Brasileiro de Pesquisas F\'{i}sicas, Rio de Janeiro, Brazil}
\affiliation{Universidade do Estado do Rio de Janeiro, Rio de Janeiro, Brazil}
\affiliation{Universidade Federal do ABC, Santo Andr\'e, Brazil}
\affiliation{University of Science and Technology of China, Hefei, People's Republic of China}
\affiliation{Universidad de los Andes, Bogot\'a, Colombia}
\affiliation{Charles University, Faculty of Mathematics and Physics, Center for Particle Physics, Prague, Czech Republic}
\affiliation{Czech Technical University in Prague, Prague, Czech Republic}
\affiliation{Center for Particle Physics, Institute of Physics, Academy of Sciences of the Czech Republic, Prague, Czech Republic}
\affiliation{Universidad San Francisco de Quito, Quito, Ecuador}
\affiliation{LPC, Universit\'e Blaise Pascal, CNRS/IN2P3, Clermont, France}
\affiliation{LPSC, Universit\'e Joseph Fourier Grenoble 1, CNRS/IN2P3, Institut National Polytechnique de Grenoble, Grenoble, France}
\affiliation{CPPM, Aix-Marseille Universit\'e, CNRS/IN2P3, Marseille, France}
\affiliation{LAL, Universit\'e Paris-Sud, CNRS/IN2P3, Orsay, France}
\affiliation{LPNHE, Universit\'es Paris VI and VII, CNRS/IN2P3, Paris, France}
\affiliation{CEA, Irfu, SPP, Saclay, France}
\affiliation{IPHC, Universit\'e de Strasbourg, CNRS/IN2P3, Strasbourg, France}
\affiliation{IPNL, Universit\'e Lyon 1, CNRS/IN2P3, Villeurbanne, France and Universit\'e de Lyon, Lyon, France}
\affiliation{III. Physikalisches Institut A, RWTH Aachen University, Aachen, Germany}
\affiliation{Physikalisches Institut, Universit\"at Freiburg, Freiburg, Germany}
\affiliation{II. Physikalisches Institut, Georg-August-Universit\"at G\"ottingen, G\"ottingen, Germany}
\affiliation{Institut f\"ur Physik, Universit\"at Mainz, Mainz, Germany}
\affiliation{Ludwig-Maximilians-Universit\"at M\"unchen, M\"unchen, Germany}
\affiliation{Fachbereich Physik, Bergische Universit\"at Wuppertal, Wuppertal, Germany}
\affiliation{Panjab University, Chandigarh, India}
\affiliation{Delhi University, Delhi, India}
\affiliation{Tata Institute of Fundamental Research, Mumbai, India}
\affiliation{University College Dublin, Dublin, Ireland}
\affiliation{Korea Detector Laboratory, Korea University, Seoul, Korea}
\affiliation{CINVESTAV, Mexico City, Mexico}
\affiliation{Nikhef, Science Park, Amsterdam, the Netherlands}
\affiliation{Radboud University Nijmegen, Nijmegen, the Netherlands}
\affiliation{Joint Institute for Nuclear Research, Dubna, Russia}
\affiliation{Institute for Theoretical and Experimental Physics, Moscow, Russia}
\affiliation{Moscow State University, Moscow, Russia}
\affiliation{Institute for High Energy Physics, Protvino, Russia}
\affiliation{Petersburg Nuclear Physics Institute, St. Petersburg, Russia}
\affiliation{Instituci\'{o} Catalana de Recerca i Estudis Avan\c{c}ats (ICREA) and Institut de F\'{i}sica d'Altes Energies (IFAE), Barcelona, Spain}
\affiliation{Uppsala University, Uppsala, Sweden}
\affiliation{Lancaster University, Lancaster LA1 4YB, United Kingdom}
\affiliation{Imperial College London, London SW7 2AZ, United Kingdom}
\affiliation{The University of Manchester, Manchester M13 9PL, United Kingdom}
\affiliation{University of Arizona, Tucson, Arizona 85721, USA}
\affiliation{University of California Riverside, Riverside, California 92521, USA}
\affiliation{Florida State University, Tallahassee, Florida 32306, USA}
\affiliation{Fermi National Accelerator Laboratory, Batavia, Illinois 60510, USA}
\affiliation{University of Illinois at Chicago, Chicago, Illinois 60607, USA}
\affiliation{Northern Illinois University, DeKalb, Illinois 60115, USA}
\affiliation{Northwestern University, Evanston, Illinois 60208, USA}
\affiliation{Indiana University, Bloomington, Indiana 47405, USA}
\affiliation{Purdue University Calumet, Hammond, Indiana 46323, USA}
\affiliation{University of Notre Dame, Notre Dame, Indiana 46556, USA}
\affiliation{Iowa State University, Ames, Iowa 50011, USA}
\affiliation{University of Kansas, Lawrence, Kansas 66045, USA}
\affiliation{Louisiana Tech University, Ruston, Louisiana 71272, USA}
\affiliation{Northeastern University, Boston, Massachusetts 02115, USA}
\affiliation{University of Michigan, Ann Arbor, Michigan 48109, USA}
\affiliation{Michigan State University, East Lansing, Michigan 48824, USA}
\affiliation{University of Mississippi, University, Mississippi 38677, USA}
\affiliation{University of Nebraska, Lincoln, Nebraska 68588, USA}
\affiliation{Rutgers University, Piscataway, New Jersey 08855, USA}
\affiliation{Princeton University, Princeton, New Jersey 08544, USA}
\affiliation{State University of New York, Buffalo, New York 14260, USA}
\affiliation{University of Rochester, Rochester, New York 14627, USA}
\affiliation{State University of New York, Stony Brook, New York 11794, USA}
\affiliation{Brookhaven National Laboratory, Upton, New York 11973, USA}
\affiliation{Langston University, Langston, Oklahoma 73050, USA}
\affiliation{University of Oklahoma, Norman, Oklahoma 73019, USA}
\affiliation{Oklahoma State University, Stillwater, Oklahoma 74078, USA}
\affiliation{Brown University, Providence, Rhode Island 02912, USA}
\affiliation{University of Texas, Arlington, Texas 76019, USA}
\affiliation{Southern Methodist University, Dallas, Texas 75275, USA}
\affiliation{Rice University, Houston, Texas 77005, USA}
\affiliation{University of Virginia, Charlottesville, Virginia 22904, USA}
\affiliation{University of Washington, Seattle, Washington 98195, USA}
\author{V.M.~Abazov} \affiliation{Joint Institute for Nuclear Research, Dubna, Russia}
\author{B.~Abbott} \affiliation{University of Oklahoma, Norman, Oklahoma 73019, USA}
\author{B.S.~Acharya} \affiliation{Tata Institute of Fundamental Research, Mumbai, India}
\author{M.~Adams} \affiliation{University of Illinois at Chicago, Chicago, Illinois 60607, USA}
\author{T.~Adams} \affiliation{Florida State University, Tallahassee, Florida 32306, USA}
\author{G.D.~Alexeev} \affiliation{Joint Institute for Nuclear Research, Dubna, Russia}
\author{G.~Alkhazov} \affiliation{Petersburg Nuclear Physics Institute, St. Petersburg, Russia}
\author{A.~Alton$^{a}$} \affiliation{University of Michigan, Ann Arbor, Michigan 48109, USA}
\author{A.~Askew} \affiliation{Florida State University, Tallahassee, Florida 32306, USA}
\author{S.~Atkins} \affiliation{Louisiana Tech University, Ruston, Louisiana 71272, USA}
% ...
%
%
\collaboration{The D0 Collaboration\footnote{with visitors from
$^{a}$Augustana College, Sioux Falls, SD, USA,
$^{b}$The University of Liverpool, Liverpool, UK,
$^{c}$UPIITA-IPN, Mexico City, Mexico,
$^{d}$DESY, Hamburg, Germany,
$^{e}$SLAC, Menlo Park, CA, USA,
$^{f}$University College London, London, UK,
$^{g}$Centro de Investigacion en Computacion - IPN, Mexico City, Mexico,
$^{h}$ECFM, Universidad Autonoma de Sinaloa, Culiac\'an, Mexico,
$^{i}$Universidade Estadual Paulista, S\~ao Paulo, Brazil,
$^{j}$Karlsruher Institut f\"ur Technologie (KIT) - Steinbuch Centre for Computing (SCC)
and
$^{k}$Office of Science, U.S. Department of Energy, Washington, D.C. 20585, USA.
}} \noaffiliation
\vskip 0.25cm

%% file: acknowledgement.tex
%
We thank the staffs at Fermilab and collaborating institutions,
and acknowledge support from the
DOE and NSF (USA);
CEA and CNRS/IN2P3 (France);
MON, NRC KI and RFBR (Russia);
CNPq, FAPERJ, FAPESP and FUNDUNESP (Brazil);
DAE and DST (India);
Colciencias (Colombia);
CONACyT (Mexico);
NRF (Korea);
FOM (The Netherlands);
STFC and the Royal Society (United Kingdom);
MSMT and GACR (Czech Republic);
BMBF and DFG (Germany);
SFI (Ireland);
The Swedish Research Council (Sweden);
and
CAS and CNSF (China).

%% file: prd_tri_submit.bbl
\begin{thebibliography}{99}
\bibitem{eta}
The pseudorapidity is given by $\eta = -\ln[\tan(\theta/2)]$, where $\theta$ is the polar angle with respect to the proton beam direction.  We define $\eta$ with respect to the $\ppbar$ interaction vertex and detector $\eta$ ($\eta_d$) with respect to the nominal center of the detector.
\end{thebibliography}
